\documentclass[%
 preprint,
  graphicx,
 amsmath,amssymb,
 aps,
]{revtex4-2}

\usepackage{acronym}
\usepackage{bm}% bold math
\usepackage{graphicx} % For including graphics
\usepackage{subcaption} % For subfigure support

\usepackage{placeins}
\usepackage{xcolor}
\usepackage{tikz}
\usepackage{amsmath}
\usepackage{amssymb}
\usepackage{array}
\usepackage{layouts}
\usepackage{xspace}

\graphicspath{{figures}}
\acrodef{TNTI}[TNTI]{turbulent/non-turbulent interface}
\acrodef{PIV}[PIV]{particle image velocimetry}
\acrodef{LIF}[LIF]{laser induced fluorescence}
\acrodef{DNS}[DNS]{direct numerical simulation}
\acrodef{TBL}[TBL]{turbulent boundary layer}
\acrodef{HWA}[HWA]{hot-wire anemometry }
\acrodef{TKE}[TKE]{turbulent kinetic energy}
\acrodef{UMZ}[UMZ]{uniform momentum zone}
\acrodef{PDF}[PDF]{probability density function}
\acrodef{CDF}[CDF]{Cumulative distribution functions}
\acrodef{UMZ}[UMZ]{Uniform Momentum Zone}

\newcommand{\fref}[1]{Fig. \ref{#1}}
\newcommand{\Fref}[1]{Fig. \ref{#1}}
\newcommand{\tref}[1]{table \ref{#1}}
\newcommand{\Tref}[1]{Table \ref{#1}}
\newcommand{\eref}[1]{equation \ref{#1}}

\newcommand{\sref}[1]{section \ref{#1}}

\definecolor{plt1}{RGB}{31, 119, 180}
\definecolor{plt2}{RGB}{255, 127, 14}
\definecolor{plt3}{RGB}{44, 160, 44}
\definecolor{plt4}{RGB}{214, 39, 40}
\definecolor{plt5}{RGB}{148, 103, 189}
\definecolor{plt6}{RGB}{140, 86, 75}
\definecolor{red}{RGB}{255, 0, 0}
\definecolor{blue}{RGB}{0, 0, 255}
\definecolor{black}{RGB}{0,0,0}

\DeclareRobustCommand\dashed{\tikz[baseline=-0.6ex]\draw[thick,dashed] (0,0)--(0.54,0); \xspace}
\DeclareRobustCommand\dotdash{\tikz[baseline=-0.6ex]\draw[thick, dash pattern=on 1pt off 2pt on 4pt off 2pt] (0,0)--(0.54,0);\xspace}

\DeclareRobustCommand\line{\tikz[baseline=-0.6ex]\draw[thick] (0,0)--(0.54,0);\xspace}

\DeclareRobustCommand\lineCircle[1]{\tikz[baseline=-0.6ex] \draw[#1,fill=#1,thick] 
    (-0.75em,0) -- (0,0) 
    (0,0) circle[radius=0.25em] 
    (0,0) -- (0.75em,0);\xspace
}

\newcommand{\lineBreakCell}[2][c]{%
  \begin{tabular}[#1]{@{}c@{}}#2\end{tabular}}
  
\begin{document}

%\preprint{}

\title{Study of the Turbulent/Non-turbulent Interface of Zero-Pressure-Gradient Turbulent Boundary Layer  Using the Uniform Momentum Zone Concept} %Title of paper



\author{Bihai Sun}
\email[]{bihai.sun@monash.edu}
\author{Callum Atkinson}
\author{Julio Soria}
%\thanks{}
\affiliation{Laboratory for Turbulence Research in Aerospace \& Combustion (LTRAC) Department of Mechanical and Aerospace Engineering, Monash University, Victoria 3800, Australia}


\date{\today}

\begin{abstract}

This paper investigates the turbulent--non-turbulent interface (TNTI) in a zero-pressure-gradient turbulent boundary layer (ZPG-TBL) using a novel, threshold-free method based on the uniform momentum zone (UMZ) concept. Requiring only planar streamwise velocity data, the method is directly applicable to experimental PIV and ensures consistent TNTI detection across simulations and experiments. Its performance is demonstrated using DNS data at $Re_\tau = 1{,}000$–$2{,}000$. The TNTI height scales with the local boundary layer thickness ($\delta$), yielding an error-function-like intermittency profile and statistics consistent with prior studies. Sensitivity to streamwise domain length is minimal. Compared to TKE- and vorticity-based methods, the UMZ-TNTI partially overlaps with the TKE interface but differs significantly from the vorticity threshold, which lies farther from the wall. Conditional averages reveal sharp velocity gradients across the TNTI, consistent with mixing-layer-like dynamics. When normalized by TNTI height and velocity jump, mean velocity profiles collapse across Reynolds numbers. Reynolds stresses respond asymmetrically: $\tilde{\overline{u'u'}}$ varies most, $\tilde{\overline{v'v'}}$ moderately, and $\tilde{\overline{w'w'}}$ least. Mean and fluctuating vorticity profiles collapse well when scaled by the UMZ-TNTI vorticity scale. A localized peak in spanwise mean vorticity is observed within the TNTI, while $\tilde{\overline{\omega_x'\omega_x'}}$ decreases across it and the other components show local maxima.

\end{abstract}

\pacs{}% insert suggested PACS numbers in braces on next line

\maketitle %\maketitle must follow title, authors, abstract and \pacs

\section{Introduction}

\Ac{TNTI} refers to the interfacial layer that separates the turbulent region of a flow from its surrounding non-turbulent region \cite{Carlos2014}. The \ac{TNTI} is closely linked to the entrainment process in turbulent flows, which is both theoretically significant and practically relevant to many engineering applications. Research on the \ac{TNTI} dates back to the 1950s, when \textcite{Corrsin1955} measured intermittency profiles in a turbulent boundary layer, a round jet, and a turbulent plane wake. The existence of a thin ``laminar superlayer'' was proposed, where viscous diffusion is the primary mechanism transmitting both mean and fluctuating vorticity across the interface.

Since then, the properties of the \ac{TNTI} have been investigated in numerous experimental and numerical studies. \textcite{Bisset2002} discovered a velocity jump at the interface using direct numerical simulation (DNS) datasets of forced and unforced wakes. Similarly, \textcite{Westerweel2009} observed a velocity jump across the interface by conditionally averaging velocity fields obtained from \ac{PIV} measurements, as well as a scalar jump in dye concentration measured by \ac{LIF}. A similar analysis of large velocity gradients was conducted by \textcite{Holzner2011} on a \ac{TBL}.

In addition to the velocity jump, the geometric properties of the interface and the mass flow rate across it have been extensively analysed. \textcite{Fiedler1966} first showed that the \ac{TNTI} surface area is much larger than its projected area, indicating that the interface is heavily folded. This result led \textcite{Townsend1980} to propose two conjectures: first, that the mass transfer per unit area is similar across different turbulent flows, with variations in entrainment rates primarily due to differences in folding intensity; and second, that large pockets of non-turbulent flow can be trapped within the turbulent region before becoming rotational, a process known as engulfment. More recent work by \textcite{deSilva2013} showed that the \ac{TNTI} surface exhibits fractal scaling, while \textcite{Borrell2016} analysed the distribution of turbulent pockets in the non-turbulent region and non-turbulent pockets within the turbulent region.

Despite extensive research on the \ac{TNTI}, the question of the most appropriate detection criterion remains unresolved\cite{Anand2009}. Early studies, limited by the velocity point measurements provided by \ac{HWA}, relied on a cutoff frequency in the streamwise velocity \cite{Corrsin1955,Fiedler1966}. Thresholding scalar concentration from \ac{LIF} measurements was also used to distinguish between turbulent and non-turbulent regions \cite{Prasad1989,Westerweel2009}. The availability of planar and three-dimensional velocity fields from \ac{PIV} and \ac{DNS} has since enabled the use of a vorticity threshold, which has become widely adopted \cite{Bisset2002,deSilva2011,Borrell2016}. Additionally, thresholds based on local \ac{TKE} fluctuations in a frame of reference moving with the freestream velocity have been applied \cite{Chauhan2014,deSilva2013}.

\Acp{UMZ} are regions with relatively uniform streamwise momentum in turbulent flows. The existence of \acp{UMZ} in turbulent boundary layers was first discovered by \textcite{Meinhart1995}, with \textcite{Adrian2000} proposing a method based on the streamwise velocity \acp{PDF} to identify \acp{UMZ} in turbulent flows. Since then, there have been extensive studies about the UMZ in both zero-pressure-gradient and adverse-pressure-gradient turbulent boundary layers \cite{Laskari2018,Senthil2020}. Some studies have suggested linking the \ac{UMZ} edge with the \ac{TNTI} \cite{deSilva2017,Sun2024,Thavamani2020}. However, due to differences between the behaviour of the \ac{TNTI} and internal \ac{UMZ} interfaces, the non-turbulent region is often neglected in \ac{UMZ} studies \cite{deSilva2016,Thavamani2020}. It has been demonstrated that a velocity jump exists across \ac{UMZ} boundaries \cite{deSilva2016}, a characteristic also observed across the \ac{TNTI}. This similarity forms the basis for using the \ac{UMZ} concept to identify the \ac{TNTI}.

This study introduces a threshold-free methodology for identifying the TNTI, addressing the limitations associated with vorticity- and TKE-based methods, which require pre-defined thresholds. Such thresholds introduce arbitrariness and human bias, complicating comparisons across datasets and flow conditions. It is also demonstrated that the optimal threshold values often exhibit Reynolds number dependence, complicating a universal application of these methods. The proposed method overcomes these challenges and provides a more consistent characterisation of the TNTI's influence on mean velocity and vorticity profiles. 

Beyond the development of the identification methodology, an extensive analysis is conducted to characterise the geometrical and statistical properties of the interface using the novel methodology presented. This includes an evaluation of interface height distributions and intermittency profiles, benchmarked against existing literature. Additionally, conditionally averaged turbulent statistics relative to the TNTI are examined to explore the flow dynamics within and around the interface.  The proposed methodology builds on these concepts to develop a TNTI identification technique that is threshold-free, experimentally compatible, and capable of detecting shear-layer-like dynamics with sharper velocity gradients as well as better collapsed conditionally averaged vorticity profiles than those typically captured by TKE-based approaches.
The interface identification procedure is described in \sref{sec:identification}. The geometric characteristics of the \ac{TNTI} are analysed in \sref{sec:geographic_properties}, followed by a sensitivity analysis in \sref{sec:sensitivity_analysis} and a comparison against other methods in \sref{sec:comparasion_methods}. Finally, conditionally averaged turbulent statistics based on \ac{TNTI} height are presented and discussed in \sref{sec:Conditional_turbulent_statistics}.

\section{Direct Numerical Simulation of Zero Pressure Gradient Turbulent Boundary Layer}
\label{sec:DNS_details}

The DNS dataset from \textcite{Sillero2013} of a ZPG TBL is used for the analysis, which is produced by the methodology developed by \textcite{Borrell2013}. The simulation code and implementation are described by \textcite{Simens2009}, with modifications introduced to achieve a higher Reynolds number \cite{Javier2013}. This DNS dataset has been thoroughly verified and used in a number of studies \cite{Sillero2013,Sillero2014}, and the readers are referred to them for a full description.

The key parameters and characteristics of the simulation are summarised in \tref{tbl:DNS_details}. The turbulent statistics from this simulation show good agreement with previous experiments and \ac{DNS} results \cite{Sillero2013,Sillero2014}. A total of 28 field snapshots were obtained from the simulation with a time separation of approximately 0.2 flow turnover times, which is sufficient to ensure the statistical independence of the samples.

\begin{table}[ht!]
\caption{Summary of important simulation parameters. In this table, $N_x$, $N_y$, and $N_z$ denote the grid size in each direction. The Taylor-microscale Reynolds number $Re_\lambda$, the Kolmogorov length $\eta$, and the Taylor microscale $\lambda$ are estimated at $y=0.6\delta$.}
\begin{ruledtabular}
\begin{tabular}{ccccccc}
$N_x$&$ N_y$&$ N_z$ & $Re_\tau$ & $Re_\lambda$ & $\delta / \eta$ & $\delta / \lambda$ \\

$15,361$&$535$&$4,096$ & $1,000$--$2,000$ & $75$--$108$ & $242$--$440$ & $14.2$--$21.4$
\end{tabular}
\end{ruledtabular}
\label{tbl:DNS_details}
\end{table}

In this paper, the $x$, $y$, and $z$ axes denote the streamwise, wall-normal, and spanwise directions, respectively. The instantaneous velocity is $\bm{u}$, with components along each axis given by $u$, $v$, and $w$. The ensemble average is represented by the overbar $\overline{\square}$. Primed variables $\cdot'$ represent fluctuating components, while capitalised variables denote mean quantities. Variables with a `$+$' superscript are normalised by the friction velocity $u_\tau$ and kinematic viscosity $\nu$. The boundary layer thickness, $\delta$, is defined as the position where the mean velocity is $99\%$ of the freestream velocity. The TNTI is denoted by $y_i$, and coordinates relative to the interface are denoted $\tilde{y}$, such that $y = y_i + \tilde{y}$. Conditional statistics with respect to the interface are denoted by a tilde over the symbol, $\tilde{\ \square\ }$.

\section{UMZ-TNTI Methodology to Identify the TNTI}
\label{sec:identification}

\begin{figure}[!htbp]

\begin{tabular}{cc}
\begin{subfigure}{0.45\textwidth}
\includegraphics[width=\textwidth]{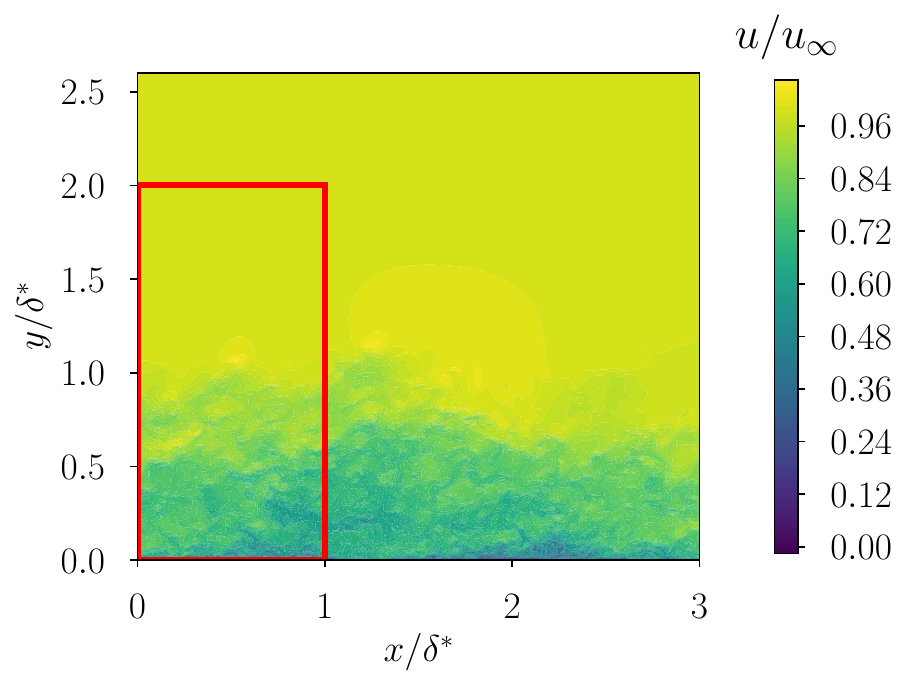} 
\caption{}
\label{fig:TNTI_identification_fig_a}
\end{subfigure}

\begin{subfigure}{0.45\textwidth}
\includegraphics[width=\textwidth]{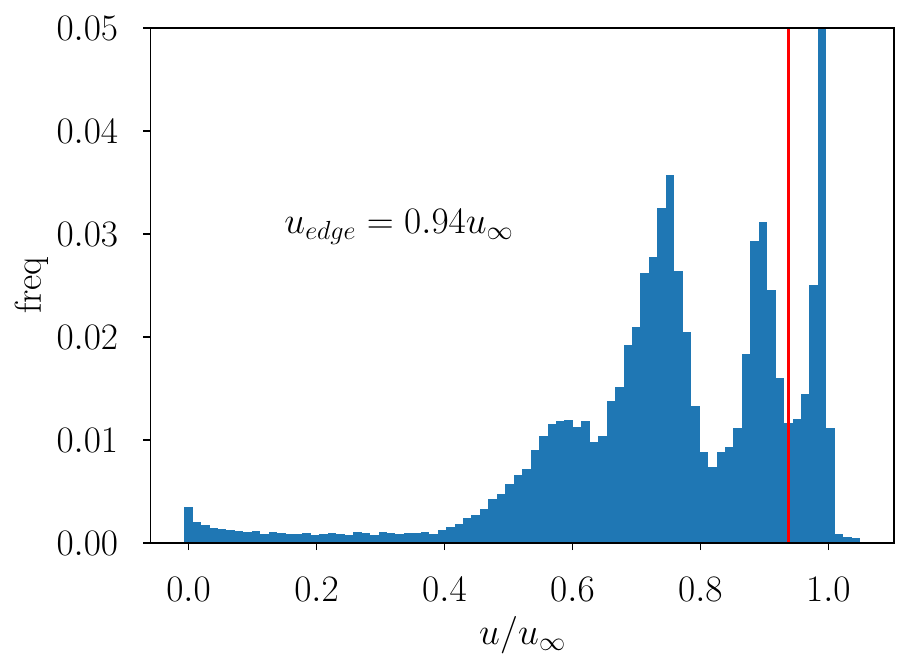} 
\caption{}
\label{fig:TNTI_identification_fig_b}
\end{subfigure} \\ 

\begin{subfigure}{0.45\textwidth}
\includegraphics[width=\textwidth]{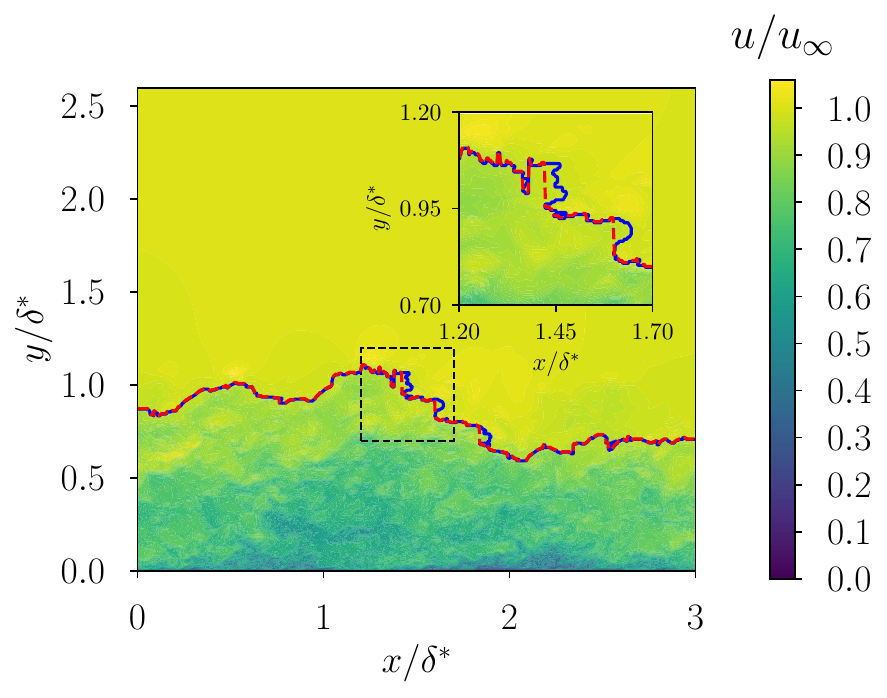} 
\caption{}
\label{fig:TNTI_identification_fig_c}
\end{subfigure}

\begin{subfigure}{0.45\textwidth}
\includegraphics[width=\textwidth]{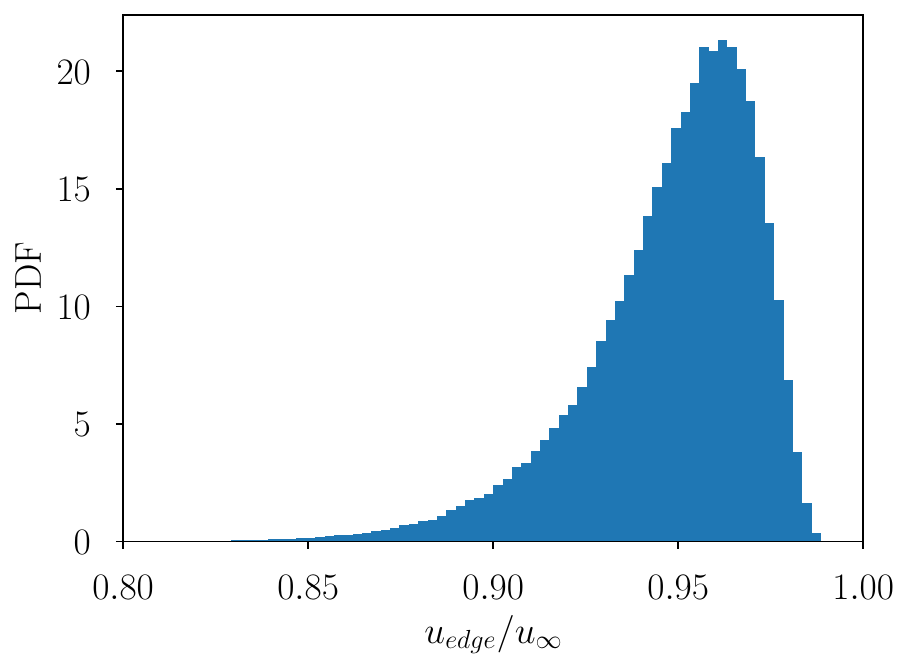} 
\caption{}
\label{fig:TNTI_identification_fig_d}
\end{subfigure}
\end{tabular}
\caption{ (a) The contour plot of an example streamwise velocity field. The red rectangle marks the sliding window at $x=0$. (b) Histogram of the velocity within the sliding window shown in (a). This histogram is clipped to a maximum frequency of 0.05. The peak representing the freestream velocity reads 0.33 and is not shown in the figure. The red vertical line marks the position of the first point of the local minimum from the freestream velocity, which is $u_{edge}$. (c) The identified \ac{TNTI} (\textcolor{blue}{\line}) and interface height ($y_{i}$, \textcolor{red}{\dashed}) overlayed on the instantaneous streamwise velocity contour. Insert: magnified plot within the black dashed square. (d)  \Ac{PDF} of the $u_{edge}$ within the entire volume. In all the plots, $\delta^*$ refers to the boundary layer thickness at the centre of the domain.}
\label{fig:TNTI_identification_fig}
\end{figure}

In the process of identifying the \ac{TNTI}, each plane in the spanwise direction of the DNS, which is periodic and therefore a homogeneous direction, is treated as statistically independent. Each realization of the flow is also assumed to be statistically independent, and all are processed individually. Therefore, this methodology is also applicable to data acquired by planar \ac{PIV}, where only statistically independent streamwise -- wall-normal velocity fields are available. For each plane, a sliding window of size $1\delta \times 2\delta$ is created and moves from the upstream direction to the downstream direction, as shown in \fref{fig:TNTI_identification_fig_a}. The boundary layer thickness in the window size refers to the thickness at the centre of the window; therefore, the sliding window grows larger as it moves downstream. As the thickness of the \ac{TNTI} scales with $\delta$, shown later in the paper, the increase in size ensures that the proportion of the turbulent and non-turbulent parts of the flow remains roughly the same throughout the process. A window height of $2\delta$ is chosen to ensure that all of the TNTI is captured and that a sufficiently large non-turbulent region is present in the window. The sensitivity of the streamwise window size on the \ac{TNTI} profile detected is discussed in \sref{sec:sensitivity_analysis}.

The histogram of the instantaneous streamwise velocity is calculated within the sliding window, as shown in \fref{fig:TNTI_identification_fig_b}. The histogram shows localised peaks corresponding to the modal velocities of each \ac{UMZ}, while the local minima represent the edge velocities that separate the \ac{UMZ}s. As expected, there is a prominent peak representing the freestream velocity. The local minima separating this peak from the rest of the turbulent flow is identified as the edge velocity $u_{edge}$, and the velocity iso-contour line corresponding to $u_{edge}$ is identified as the \ac{TNTI} within the window. As the window moves downstream, new values of $u_{edge}$ are determined by the same process, and the \ac{TNTI} location is updated accordingly.
The identified velocity contour is not a single curve, instead there are pockets of turbulent region in the non-turbulent part of the flow, as well as non-turbulent pockets in the turbulent part of the flow. The edges of these pockets are closed and clearly do not form part of the TNTI. Therefore, the boundaries of these pockets are removed from the identified interface before the subsequent analysis.

An example of the \ac{TNTI} found using this methodology is shown in \fref{fig:TNTI_identification_fig_c}. The \ac{TNTI} is clearly not smooth, and its height varies significantly over the domain of $2\delta$, with some streamwise locations extending beyond the local boundary layer thickness. The interface also folds back onto itself in several positions, leading to multiple \ac{TNTI} locations at a particular streamwise position. 
%Large-scale indentations coexist with smaller pockets, and there are turbulent bubbles in the non-turbulent region as well as non-turbulent bubbles in the turbulent region. These bubbles in the 2D slices are not necessarily closed in the third dimension but may be a part of spanwise engulfment. 

In this paper, the lower envelope of the interface is considered for the interface height $y_{i}$, as indicated by the red dashed line in \fref{fig:TNTI_identification_fig_c}. This ensures that the entire region below the interface is fully turbulent.

\Fref{fig:TNTI_identification_fig_d} presents the \ac{PDF} of $u_{edge}/u_\infty$ within the computational domain for one flow field. The distribution is negatively skewed, with a tail extending towards lower $u_{edge}$. The most probable $u_{edge}$ is $0.957\, u_\infty$, while the mean $u_{edge}$ is $0.950\, u_\infty$.

\FloatBarrier
\section{Geometric properties of the TNTI}
\label{sec:geographic_properties}

Although the presented methodology can be applied to the entire velocity field, the analysis presented in this paper is limited to three streamwise stations to isolate and study the effect of Reynolds number on the \ac{TNTI} properties. Each station contains the full domain in the $z$ and $y$ directions, and the extents in the $x$ direction are summarised in \tref{tbl:station_extends}.

\begin{table}[ht!]
\caption{ Streamwise extents of the analysis domains. These stations contain the full extent in $y$ and $z$ directions. The starred symbols ($\cdot^*$) represent the physical quantities at the centre of the corresponding station.}
\begin{ruledtabular}
\begin{tabular}{c|ccccc}
Station & $Re_\tau^*$ & $Re_\theta^*$ & $L_x/\delta^*$ \\ \hline
1       & 1,045                & 3,054                  & 3            \\ \hline
2       & 1,430                & 4,521                  & 3            \\ \hline
3       & 1,965                & 6,446                  & 3            \\
\end{tabular}
\end{ruledtabular}
\label{tbl:station_extends}
\end{table}

The mean TNTI height and the standard deviation of the TNTI height were compared with previously published results in \tref{tbl:TNTI_stat_compare}. \Tref{tbl:TNTI_stat_compare} shows a wide range of mean TNTI heights and standard deviations reported from the existing literature, which do not show a strong Reynolds number dependency. The results by \textcite{Jimenez2010} and \textcite{Eisma2015} have higher mean TNTI heights because a threshold in vorticity is used to identify the interface. Excluding these two results, $\overline{y_{i}} / \delta$ and $\sigma\left(y_i\right) / \delta$ obtained in this study are comparable with values reported in the literature.

\begin{table}[!htbp]
\caption{ Comparison of mean position and standard deviation of the TNTI with existing literature. }
\begin{ruledtabular}
\begin{tabular}{lcccc} 
& $R e_\tau$ & $\overline{y_i} / \delta$ & $\sigma\left(y_i\right) / \delta$ & $\sigma\left(y_i\right) / \overline{y_i}$ \\
\textcite{Jimenez2010} & 692 & 0.92 & 0.10 & 0.11\\
\textcite{Chen1978} & 1,190 & 0.82 & 0.13 & 0.16 \\
\textcite{Kovasznay1970} & 1,240 & 0.78 & 0.14 & 0.18 \\
\textcite{Corrsin1955} & $<2,000$ & 0.80 & 0.16 & 0.20 \\
\textcite{Eisma2015}   & 2,053 & 0.90 &0.18 &0.20\\
\textcite{Hedley1974} & 5,100 & 0.75 & 0.24 & 0.32 \\
\textcite{Chauhan2014} & 14,500 & 0.82 & 0.13 & 0.17\\
Current study station 1 & 1,045 & 0.74 & 0.17 & 0.23\\
Current study station 2 & 1,430 & 0.75 & 0.16 & 0.21\\
Current study station 3 & 1,965 & 0.77 & 0.16 & 0.21\\
\end{tabular}
\end{ruledtabular}
\label{tbl:TNTI_stat_compare}
\end{table}

%The \ac{PDF}s of the instantaneous \ac{TNTI} height $y_i$, normalised by the local boundary layer thickness, are plotted in \fref{fig:TNTI_geometics_fig_a} for three stations. The \ac{PDF}s are calculated by combining all the data points within each station. Although the distribution of $u_{edge}$ is negatively skewed, the \ac{PDF} of the normalised interface height exhibits a normal distribution for all three stations. \Fref{fig:TNTI_geometics_fig_b} shows the intermittency profile of the \ac{TNTI}. Intermittency refers to the proportion of time for which the velocity is turbulent at a given location\cite{Corrsin1943}. In this paper, the intermittency profiles were calculated by assigning the instantaneous turbulent region a value of one and the non-turbulent region a value of zero, then averaging across the $x$ and $z$ directions as well as statistically independent samples. The intermittency profile is also equivalent to the \ac{CDF} of $y_i$. Since $y_i$ follows a normal distribution, the intermittency profiles take the form of error functions. Again, the intermittency profiles from the three stations are very similar to each other. The intermittency profile also shows that the TNTI location varies a lot within the turbulent boundary layer, with 2\% of the time higher than the boundary layer thickness and about 2\% of the time lower than half boundary layer height. 

\begin{figure}[!htbp]
\centering
\begin{tabular}{cc}
\begin{subfigure}{0.49\textwidth}
\includegraphics[width=\textwidth]{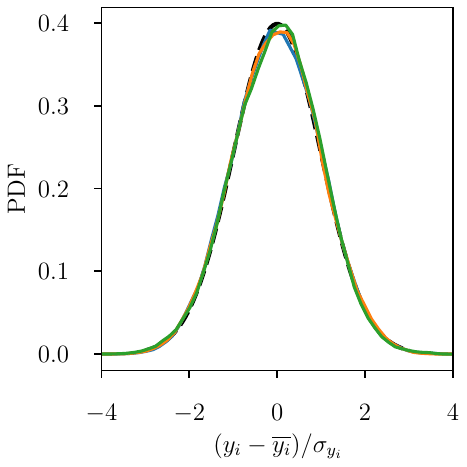} 
\caption{}
\label{fig:TNTI_geometics_fig_a}
\end{subfigure}

\begin{subfigure}{0.49\textwidth}
\includegraphics[width=\textwidth]{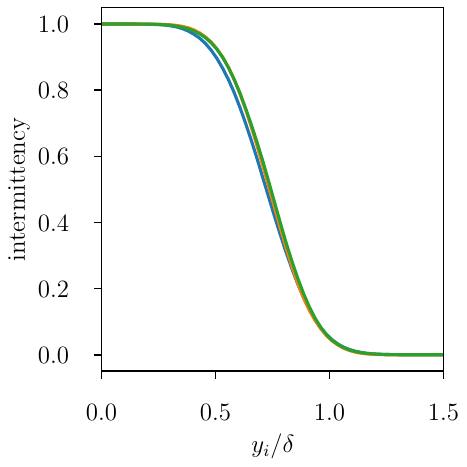} 
\caption{}
\label{fig:TNTI_geometics_fig_b}
\end{subfigure}
\end{tabular}
\caption{ (a) Disribution of the \ac{TNTI} height normalised by mean interface height and standard deviation. (b) The intermittency profiles of the interface. \dashed, standard normal distribution;  \textcolor{plt1}{\line} Station 1, \textcolor{plt2}{\line} Station 2,  \textcolor{plt3}{\line} Station 3.}
\label{fig:TNTI_geometics_fig}
\end{figure}

The \acp{PDF} of the instantaneous \ac{TNTI} height, $y_i$, normalized by the local boundary layer thickness, are shown in \fref{fig:TNTI_geometics_fig_a} for three streamwise stations. Each \ac{PDF} is computed by aggregating all data points within the corresponding station. Although the distribution of $u_{edge}$ is negatively skewed, the \acp{PDF} of the normalized interface height exhibit a nearly Gaussian distribution across all stations.

\Fref{fig:TNTI_geometics_fig_b} presents the intermittency profiles of the \ac{TNTI}. Intermittency is defined as the proportion of time that the velocity at a given location is classified as turbulent~\cite{Corrsin1943}. In this study, the intermittency is computed by assigning a value of one to turbulent regions and zero to non-turbulent regions, followed by averaging over the $x$ and $z$ directions and across statistically independent samples. 

The intermittency profiles presented in \fref{fig:TNTI_geometics_fig_b} also correspond to the cumulative distribution functions (CDFs) of the interface height, $y_i$. Since $y_i$ is approximately normally distributed, as presented in Figure 2a, the CDFs resemble error functions. The profiles at all three stations are closely aligned, suggesting consistent statistical behavior of the turbulent/non-turbulent interface (TNTI) along the streamwise direction. These profiles also demonstrate that the TNTI exhibits significant fluctuations, as the intermittency profile is above zero at the boundary layer edge, where $y_i/\delta = 1$, as well as in the middle of the boundary layer, where $y_i/\delta = 0.5$. This behavior is consistent with experimental observations by \citet{Chauhan2014}, who used a TKE-based method and reported similar intermittency trends, despite differences in dataset and scaling.

\section{Sensitivity analysis of the UMZ streamwise window size on the TNTI}
\label{sec:sensitivity_analysis}
Although the UMZ-TNTI methodology does not require a threshold to identify the TNTI, one parameter must be predetermined before performing the analysis: the streamwise domain size used to calculate the velocity PDF for UMZ identification, $L_x$. Previous studies have shown that the number of \ac{UMZ} found in the flow depends on the streamwise extent of the analysis domain\cite{Thavamani2020,deSilva2016}, therefore there is a possibility that the \ac{TNTI} identified also depends on the streamwise extent of the analysis.

\begin{figure}[!htbp]
\centering
\begin{tabular}{cc}
\begin{subfigure}{0.49\textwidth}
\includegraphics[width=\textwidth]{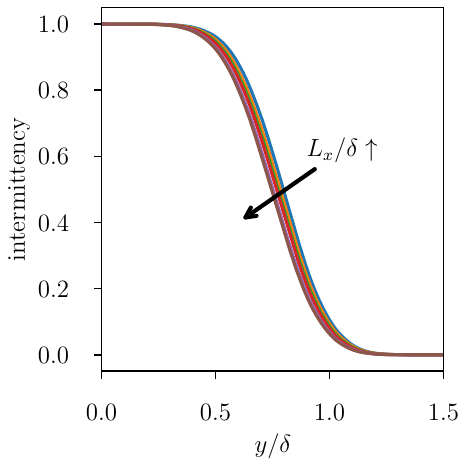} 
\caption{}
\label{fig:diff_window_size_diff_case_station1}
\end{subfigure}
\\
\begin{subfigure}{6.69 in}
\includegraphics[width=\textwidth]{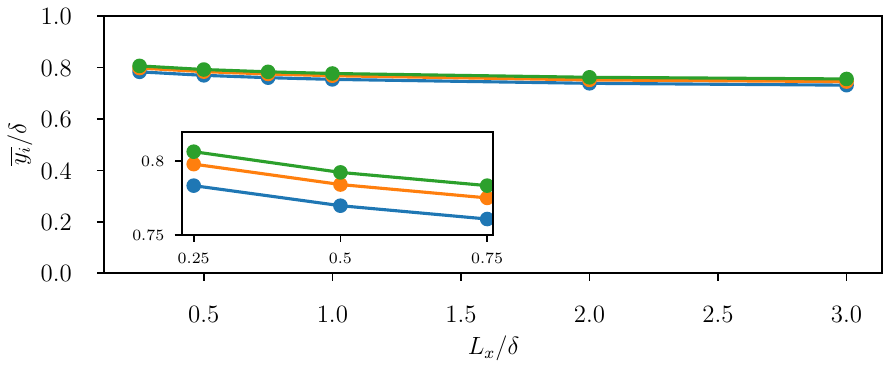} 
\caption{}
\label{fig:diff_window_size_mean_height}
\end{subfigure}
\end{tabular}
\caption{ (a) The intermittency profiles of the \ac{TNTI} found in station 1 using different streamwise domain length ($L_x$). Stations 2 and 3 (not shown) follow the same trend. \textcolor{plt1}{\line} $L_x/\delta=0.25$, \textcolor{plt2}{\line} $L_x/\delta=0.5$, \textcolor{plt3}{\line} $L_x/\delta=0.75$, \textcolor{plt4}{\line} $L_x/\delta=1.0$, \textcolor{plt5}{\line} $L_x/\delta=2.0$, \textcolor{plt6}{\line} $L_x/\delta=3.0$. (b) Mean \ac{TNTI} interface height found in three stations using different domain lengths. Insert: magnified plot of $L_x/\delta$ between 0.25 and 0.75. \lineCircle{plt1} Station 1, \lineCircle{plt2} Station 2, \lineCircle{plt3} Station 3. }
\label{fig:sensitivity_analysis}
\end{figure}

In this section, six different streamwise domain sizes are considered, which are $0.25\delta$, $0.5\delta$, $0.75\delta$, $1\delta$, $2\delta$ and $3\delta$. Because the height of the \ac{TNTI} scales with $\delta$, the window sizes are chosen in terms of $\delta$ instead of viscous units, as suggested by \textcite{deSilva2016}. The resulting intermittency profiles are summarised in \fref{fig:sensitivity_analysis}.

\Fref{fig:diff_window_size_diff_case_station1} shows a small variation of the intermittency profiles with streamwise domain size. As the streamwise domain size increases, there are a visible but minor changes in the intermenttency profile. This can be attributed to the fact that the domain size introduces an implicit smoothing filter to the \ac{PDF} of the streamwise velocity, and the \ac{UMZ} that includes the freestream is slightly thicker when a longer domain is used.

Even though there is a slight decrease in the mean TNTI height, \Fref{fig:diff_window_size_mean_height} shows that these changes are minimal for the window size range of 0.25$\delta$ to 3$\delta$. While the effect of the sliding window size on TNTI statistics could, in principle, be influenced by Reynolds number, our results indicate that for the range of Reynolds numbers studied, this influence is minimal. As shown in \fref{fig:diff_window_size_mean_height}, the intermittency profiles exhibit negligible variation across Reynolds numbers, suggesting that the observed statistical behavior is robust with respect to both Reynolds number and window size within the investigated parameter space.
The aspect ratio of \fref{fig:diff_window_size_mean_height} has been selected to reflect the physical scaling, as both axes are normalized by the boundary layer thickness ($\delta$), illustrating that the fluctuations in TNTI height are small compared to the domain length

\section{Comparison of UMZ-TNTI with other methodologies}
\label{sec:comparasion_methods}

In this section, the \ac{TNTI} measured using the UMZ-TNTI method is compared with the \ac{TNTI} measured using the local \ac{TKE} method\cite{Chauhan2014} and the vorticity threshold method\cite{Javier2013}.

A local turbulent kinetic energy in the frame of reference moving with $u_\infty$, over a $3 \times 3$ grid, is defined as:

\begin{equation}
k_{2C}=\frac{100}{9 u_{\infty}^2} \sum_{i,j=-1}^1\left[\left(u_{i,j}-u_{\infty}\right)^2+\left(v_{i,j}\right)^2\right].
\end{equation}

\noindent As the analysis by \textcite{Chauhan2014} was performed on 2C–2D PIV data, the local turbulent kinetic energy is defined only in a two-dimensional plane using two velocity components. However, because the full definition of the turbulent kinetic energy contains all three components, this definition is incomplete. 
As volumetric 3C–3D velocity fields from \ac{DNS} are available for this study, the definition of local turbulent kinetic energy is expanded to the third component over the same two-dimensional plane as:

\begin{equation}
k_{3C}=\frac{100}{9 u_{\infty}^2} \sum_{i,j=-1}^1\left[\left(u_{i,j,k}-u_{\infty}\right)^2+\left(v_{i,j,k}\right)^2+\left(w_{i,j,k}\right)^2\right].
\end{equation}

\noindent In the non-turbulent region, $k$ is nearly the same as the freestream turbulence intensity $\overline{u'u'}/u_\infty^2$, whereas in the turbulent region, $k$ is much higher. Therefore, a properly selected threshold $k_\text{th}$ can be used to separate the turbulent region from the non-turbulent region. However, as the freestream turbulence intensity differs between \ac{DNS} and experimental setups, and $k$ in the turbulent region is Reynolds number dependent, $k_\text{th}$ must be determined on a case-by-case basis. 

Two criteria are suggested to find a suitable $k_\text{th}$: first, the intermittency profile resulting from $k_\text{th}$ should agree with an error function; second, the mean and standard deviation of $y_i$ should satisfy $\overline{y_i} + 3\sigma\left(y_i\right) \approx \delta'$. Note that the boundary layer thickness defined by \textcite{Chauhan2014} ($\delta'$) is not based on 99\% of the freestream velocity. For the boundary layer thickness definition used in this paper, the second criterion is equivalent to $\overline{y_i} + 3\sigma\left(y_i\right) \approx 1.22\delta$. The threshold-finding process is implemented as an optimisation routine and then applied to $k_{2C}$, yielding $k_{2C,th}$, and to $k_{3C}$, yielding $k_{3C,th}$. The resulting values of $k_{2C,th}$ and $k_{3C,th}$ for the three stations are summarised in \tref{tbl:k_th}.

\begin{table}[!htbp]
\caption{ $k_{th}$ found for the 2-component and 3-component definition of $k$ from the current dataset compared with $k_{th}$ found by \textcite{Chauhan2014}.}
\begin{ruledtabular}
\begin{tabular}{c|cccc}
                    & Station 1 & Station 2 & Station 3 & \lineBreakCell{\textcite{Chauhan2014}\\$Re_\tau \approx 14,500$} \\ \hline
$k_{2C,th}$ &0.18 &0.16 &0.13 &0.12 \\ \hline
$k_{3C,th}$ &0.25 &0.23 &0.20 & -  \\
\end{tabular}
\end{ruledtabular}
\label{tbl:k_th}
\end{table}

The table shows a strong Reynolds number dependency for both $k_{2C,th}$ and $k_{3C,th}$, with a 28\% difference across the three stations for $k_{2C,th}$ and a 20\% difference for $k_{3C,th}$. Both thresholds decrease with increasing Reynolds number, showing that $k$ grows slower in the wall-normal direction compared with $\delta$. Therefore, at higher Reynolds numbers, a smaller threshold is needed so that the \ac{TNTI} height scales with $\delta$. If $k_{2C,th}$ is extrapolated to $Re_\tau = 14,500$, as reported by \textcite{Chauhan2014}, $k_{2C,th}$ would be much smaller. This can be attributed to differences in the freestream turbulence level between the two datasets, as well as measurement uncertainty in the experiment. For each station, $k_{3C,th}$ is higher than the corresponding $k_{2C,th}$ because of the additional spanwise velocity fluctuations included in the calculation; otherwise, the same trend is observed.

This result shows that it is inappropriate to use a single $k_{th}$ to identify \ac{TNTI} for flows at different Reynolds numbers. It is acceptable for the analysis by \textcite{Chauhan2014} because of the limited streamwise domain, as well as the high Reynolds number leading to a small boundary layer growth rate. However, for the lower Reynolds numbers, such as the ones used in this study, it is essential to use different thresholds for different stations. 

\begin{figure}[!htbp]
\centering
\begin{tabular}{ccc}
\begin{subfigure}{0.32\textwidth}
\includegraphics[width=\textwidth]{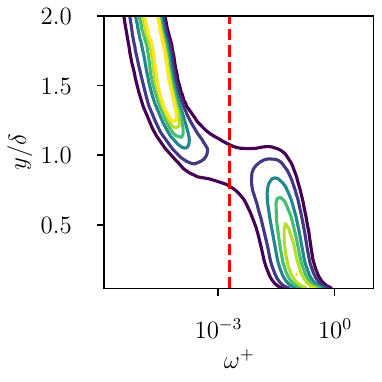} 
\caption{}
\label{fig:JPDF_omega1}
\end{subfigure}

\begin{subfigure}{0.32\textwidth}
\includegraphics[width=\textwidth]{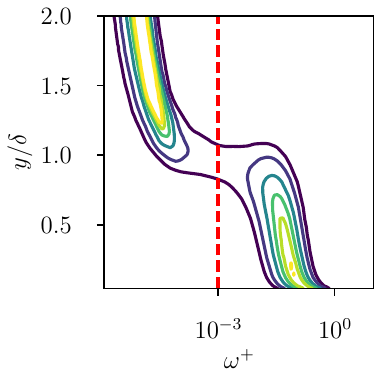} 
\caption{}
\label{fig:JPDF_omega2}
\end{subfigure}

\begin{subfigure}{0.32\textwidth}
\includegraphics[width=\textwidth]{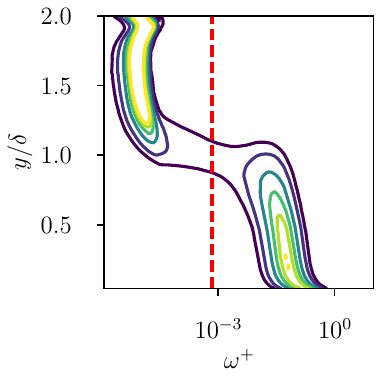} 
\caption{}
\label{fig:JPDF_omega3}
\end{subfigure}
\end{tabular}
\caption{ Premultiplied Joint PDF of $\omega^+$ and $y/\delta$ for (a) station 1, (b) station 2 and (c) station 3. The contour levels corresponds to $99\%$, $80\%$, $60\%$, $40\%$, $20\%$ and $10\%$ of the joint CDF velue. Red verticle lines represent the threshold values.}
\label{fig:JPDF_omega}
\end{figure}

The other methodology compared with the presented new approach is the imposition of a threshold on the total vorticity ($\omega$) of the flow\cite{Bisset2002}. An appropriate threshold $\omega_{th}^+$ can be found by analysing the joint PDF of $\omega^+$ and $y/\delta$\cite{Borrell2016}. For the three stations, these PDFs are presented in \fref{fig:JPDF_omega}. The figure shows a separation of scales for $\omega^+$, with two distinct peaks in the PDF connected by a plateau, where the saddle point marks the desired $\omega_{th}^+$. For all three stations, the plateau sits along $y/\delta=1$ and extends across a wide range of vorticity values of more than two orders of magnitude in $\omega^+$, while maintaining a vertical thickness of approximately $0.2\delta$. The size and flatness of this region lead to high uncertainty in determining a precise threshold, and the properties of the TNTI vary significantly within this plateau zone \cite{Borrell2016}.

In this paper, the thresholds are determined by selecting the middle point between the two peaks of the PDF in the log-linear scale. The \(\omega_{th}^+\) found for the three cases are summarised in \tref{tbl:Omega_th}. \textcite{Borrell2016} also suggested another way to non-dimensionalise vorticity so that the statistics are better converged across different Reynolds numbers,

\begin{equation}
\omega^\star=\omega Re_\tau^{1 / 2}\left(\nu / u_\tau^2\right),
\label{equ:vorticity_star_notation}
\end{equation}

\noindent and it is compared with \(\omega_{th}^\star\) found in this paper. \Tref{tbl:Omega_th} shows a strong relationship between the Reynolds number and \(\omega_{th}^+\), and a relatively weaker correlation between the Reynolds number and \(\omega_{th}^\star\). The value of \(\omega_{th}^\star\) reported by \textcite{Borrell2016} is comparable to that at the third station in this study. \Fref{fig:JPDF_omega} indicates that the positions and shapes of the turbulent peaks in the PDFs are nearly identical, and the variation in \(\omega_{th}^\star\) across the three stations can be attributed more to the changes in the non-turbulent peak. However, since non-turbulent peaks are more susceptible to numerical artefacts, this selection criterion is not entirely physical. A similar conclusion was drawn by \textcite{Borrell2016}.

\begin{table}[!htbp]
\caption{ \(\omega_{th}^\star\) and \(\omega_{th}^+\) found from the current dataset compared with \textcite{Borrell2016}.}
\begin{ruledtabular}
\begin{tabular}{c|cccc}
                    & Station 1 & Station 2 & Station 3 & \textcite{Borrell2016} \\ \hline
\(\omega_{th}^+\) &\(2.00\times 10^{-3 } \)&\(1.00\times 10^{-3 } \)&\(5.00\times 10^{-4 } \)& - \\ \hline
\(\omega_{th}^\star\) &\(6.47\times 10^{-2 } \)&\(3.78\times 10^{-2 } \)&\(2.22\times 10^{-2 } \)&\(2.20\times 10^{-2 } \) \\
\end{tabular}
\end{ruledtabular}
\label{tbl:Omega_th}
\end{table}

\begin{figure}[!htbp]
\centering
\begin{tabular}{cc}
\begin{subfigure}{0.49\textwidth}
\includegraphics[width=\textwidth]{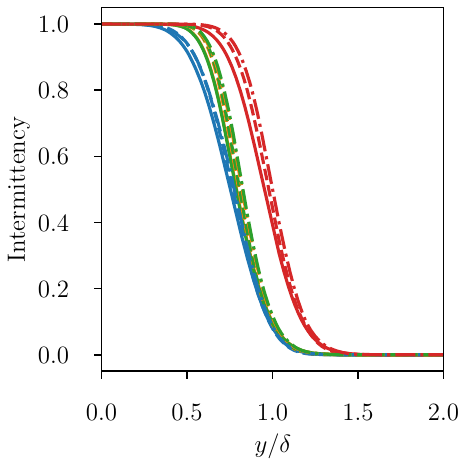} 
\caption{}
\label{fig:compare_method_intermittency}
\end{subfigure}

\begin{subfigure}{0.49\textwidth}
\includegraphics[width=\textwidth]{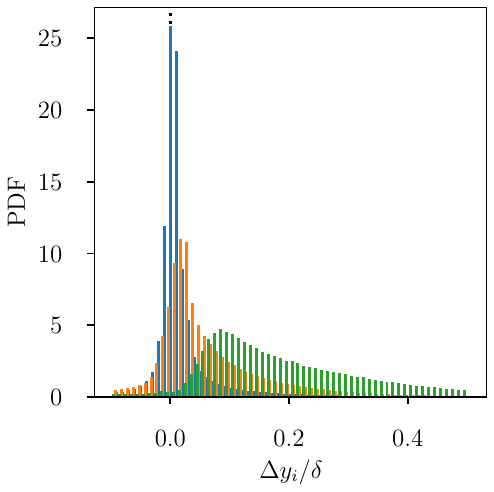} 
\caption{}
\label{fig:compare_method_interface}
\end{subfigure}
\end{tabular}
\caption{ (a) The intermittency profiles of the \ac{TNTI} found using different methods. \dotdash Station 1, \dashed Station 2, \line Station 3; \textcolor{plt1}{\line}  proposed UMZ method, \textcolor{plt2}{\line} local TKE method using 2C definition of $k$, \textcolor{plt3}{\line} local TKE method using 3C definition of $k$, \textcolor{plt4}{\line} vorticity threshold method. (b) PDF of the difference in interface height between: \textcolor{plt1}{$\blacksquare$} local TKE with 2C and 3C definition of $k$, \textcolor{plt2}{$\blacksquare$} local TKE using 2C definition of $k$ with the UMZ-TNTI method, and  \textcolor{plt3}{$\blacksquare$} vorticity threshold method with the UMZ-TNTI method.}
\label{fig:compare_method}
\end{figure}

The intermittency profiles calculated from the local TKE method, as well as the vorticity method, are compared with the UMZ-TNTI method in \fref{fig:compare_method_intermittency}. \Fref{fig:compare_method_intermittency} shows that all of the intermittency profiles behave like an error function. The intermittency profiles from the 2C and 3C local TKE methods are identical; therefore, the two methods are equivalent except for the difference in threshold value. The interface identified by the local TKE method has a slightly higher mean than the UMZ-TNTI method using UMZ, as well as a slightly smaller standard deviation. The interface identified using the vorticity threshold has a higher mean than the other methods, and the standard deviation is comparable with the profile from the presented method. Comparing the profiles from different stations, the TNTI identified using the vorticity threshold has the most variation, suggesting that the interface does not scale with \(\delta\) for different Reynolds numbers. The TNTI identified using the UMZ method has the least variation; therefore, it scales well with \(\delta\). For all methods, the intermittency profile shows more Reynolds number dependence closer to the wall than near the freestream.

%In addition, the \ac{TNTI} identified using different methods on the same representative instantaneous velocity fields is shown in \fref{fig:compare_method_interface}. This figure shows that the interfaces found using the 2C and 3C local TKE methods overlap each other; therefore, the same interface is identified, which agrees with \fref{fig:compare_method_intermittency}. The interfaces found using the local TKE method and the UMZ method overlap for part of the flow, for example, around \(x/\delta^*\approx1\), and are different in other parts. The interface found by the vorticity threshold does not overlap with other methods at all, and the geometry is very different, with a larger portion of the interface folding back on itself.

In order to give a more detailed view of the relative positions of the TNTI found by different methodologies, the distributions of the difference in the height, $\Delta y_i$, are plotted as PDFs in \fref{fig:compare_method_interface}. Three comparison are presented in \fref{fig:compare_method_interface}, which are the distribution of differences in TNTI height between, first, local TKE method using 2C verses 3C definition of $k$, second, local TKE method with $k_{2C}$ verses UMZ-TNTI method, and third, and vorticity threshold verse UMZ-TNTI method. 

The figure shows that the difference between the interface identified using the local TKE method with 2C and 3C definitions of $k$ differs less than $0.02\delta$ in height. As shown in the later section \sref{sec:Conditional_turbulent_statistics}, this difference is less than the interface thickness, $\delta_\omega$, therefore the two TNTI interfaces are indistinguishable. The result agrees with the intermittency profiles shown in \fref{fig:compare_method_intermittency}, and therefore, in the following discussion, the two interfaces are not differentiated. \fref{fig:compare_method_interface} also shows that a large portion of the TNTI found by the local TKE method is close to the TNTI identified using the UMZ-TNTI method, and where they are further apart, the TNTI identified using the local TKE method is higher. It also shows that the interface found by the vorticity threshold rarely overlaps with the interface using other methods and is, on average, 0.18$\delta$ further away from the wall than that identified using the UMZ-TNTI method.

\FloatBarrier
\section{Conditionally averaged turbulent statistics about the interface}
\label{sec:Conditional_turbulent_statistics}
The conditional turbulent statistics are calculated by following this process. First, for each $x$ and $z$ location, the instantaneous velocity and vorticity are interpolated to a grid $\tilde{y}$ originating at the interface, with a grid spacing equal to the average spacing of $y$ at the mean \ac{TNTI} height, and in the range of $-0.1\delta$ to $0.1\delta$. $\tilde{y}$ has the same positive direction as $y$, therefore negative $\tilde{y}$ corresponds to the turbulent region below the \ac{TNTI}, and positive $\tilde{y}$ corresponds to the non-turbulent region above the \ac{TNTI}. Then, an ensemble average is performed on the interpolated dataset for the mean streamwise and wall-normal velocities about the \ac{TNTI} ($\tilde{U}$,  $\tilde{V}$), Reynolds stresses ($\tilde{\overline{u'u'}}$, $\tilde{\overline{v'v'}}$, $\tilde{\overline{w'w'}}$, $\tilde{\overline{u'v'}}$), mean spanwise vorticity ($\tilde{\Omega_z}$) and vorticity fluctuation profiles ($\tilde{\overline{\omega_x\omega_x}}$, $\tilde{\overline{\omega_y\omega_y}}$, $\tilde{\overline{\omega_z\omega_z}}$).

\subsection{Conditionally averaged mean velocity profiles}
\label{sec:Conditioned_mean_velocity_profiles}
\begin{figure}[!htbp]
\captionsetup[subfigure]{aboveskip=-1pt,belowskip=-1pt}
\centering
\begin{tabular}{cc}
\begin{subfigure}{0.49\textwidth}
\includegraphics[width=\textwidth]{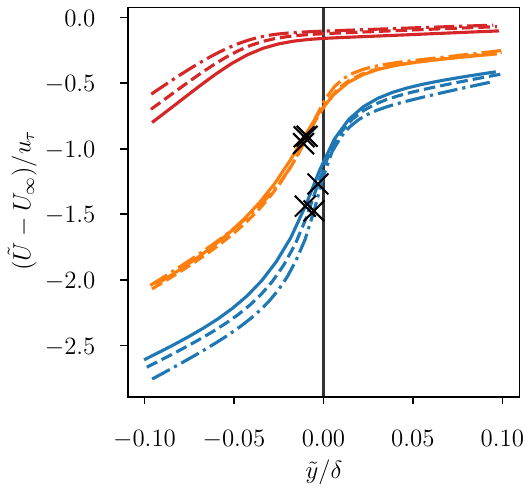} 
\caption{}
\label{fig:mean_u}
\end{subfigure}

\begin{subfigure}{0.49\textwidth}
\includegraphics[width=\textwidth]{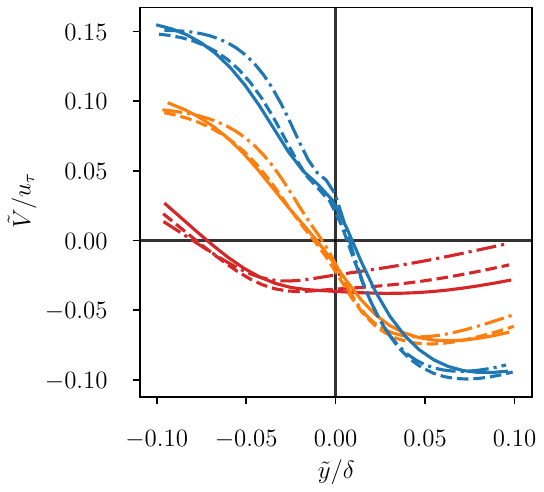} 
\caption{}
\label{fig:mean_v}
\end{subfigure}
\end{tabular}
\vspace{-1.5em}
\caption{Conditionally averaged mean velocities with respect to the \ac{TNTI}: (a) Streamwise velocity, $\tilde{U}$; (b) Wall-normal velocity, $\tilde{V}$.  \textcolor{plt1}{\line}  UMZ-TNTI method, \textcolor{plt2}{\line} Local TKE method using 2C definition of $\tilde{k}$, \textcolor{plt4}{\line} Vorticity threshold method, \dotdash Station 1, \dashed Station 2, \line Station 3. $\times$ marks the position of the largest $\tilde{U}$ gradient in (a).}

\label{fig:cond_stat_mean}
\end{figure}

Conditionally averaged mean streamwise velocities are plotted in \Fref{fig:mean_u}. For the mean streamwise velocity, the profiles found by both the local TKE method and the UMZ-TNTI method show a narrow region of sharp increase in velocity, with the position of the highest velocity gradient close to the interface. The same profile found by the vorticity threshold method shows a monotonic increase below the interface and nearly constant above the interface. Because the \ac{TNTI} resides in the wake region of the turbulent boundary layer, where the outer velocity deficit scaling applies, \Fref{fig:mean_u} plots the $\tilde{U}$ profiles in deficit form. The figure shows that the profile found by the local TKE method shows the best convergence across different Reynolds numbers, while more spread can be observed for the other two methods. 

$\tilde{V}$ profiles resulting from the three \ac{TNTI} identification methods are shown in \Fref{fig:mean_v}. A sharp decrease in $\tilde{V}$ can be observed in the profiles found by the local TKE method and the UMZ-TNTI method. Compared to $\tilde{U}$, the region of sharp decrease is wider, with the profiles levelling at $\pm0.05\delta$. The $\tilde{V}$ profiles found by the vorticity threshold method are relatively constant across the \ac{TNTI}, with a maximum velocity of less than $0.05u_\tau$. For all profiles, $\tilde{V}$ at the interface is roughly one order of magnitude smaller than $\tilde{U}$. The sharp velocity gradients near the interface observed in our TKE- and UMZ-based profiles are consistent with experimental findings reported by \citet{Chauhan2014}, who also observed a similarly narrow region of rapid change in streamwise velocity.

\begin{figure}[!htbp]

\includegraphics[width=4 in]{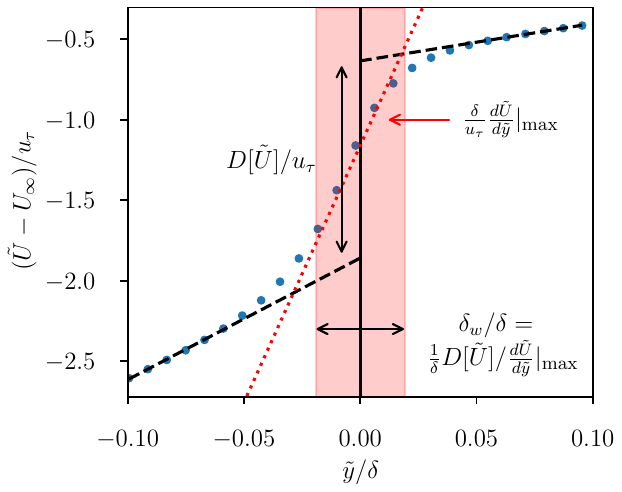} 

\caption{ Velocity, vorticity and length scales identified from the $\tilde{U}$ profile: difference in mean streamwise velocity across the interface ($D[\tilde{U}]$), maximum mean velocity gradient ($\frac{d \tilde{U}}{d\tilde{y}}|_{\mbox{\small{max}}}$), and interface thickness ($\delta_w$).  }
\label{fig:TNTI_schematics}
\end{figure}

The conditionally averaged mean velocity profiles suggest that the flow within the \ac{TNTI} found by the UMZ-TNTI method and the local TKE method behaves like a mixing layer with linearly changing velocities on either side of the mixing layer. This behaviour suggests the velocity, length, and vorticity scales within the interface, as shown in \Fref{fig:TNTI_schematics}. The mean streamwise velocity change across the interface, $D[\tilde{U}]$, can be found by extrapolating the linear velocity profiles above and below the interface to $\tilde{y}=0$, then taking the difference of the interpolated value. From $D[\tilde{U}]$ and the maximum velocity gradient within the interface $\frac{d \tilde{U}}{d\tilde{y}}|_{\mbox{\small{max}}}$, a vorticity thickness of the \ac{TNTI}, $\delta_w$, can be defined as

\begin{equation}
\delta_w= \frac{D[\tilde{U}]}{\frac{d \tilde{U}}{d\tilde{y}}|_{\mbox{\small{max}}}}.
\end{equation}

\begin{figure}[!htbp]

\includegraphics[width=3.37 in]{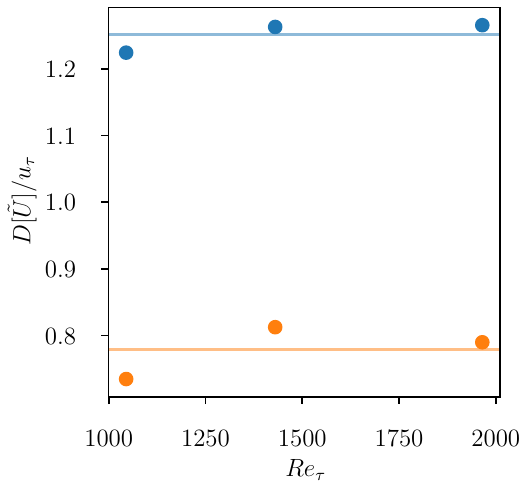} 

\caption{ The velocity jump across the TNTI, $D[\tilde{U}]$, normalised by the viscous velocity. \textcolor{plt1}{$\bullet$} UMZ-TNTI method, \textcolor{plt2}{$\bullet$} Local TKE method. Faint coloured lines represent the averaged $D[\tilde{U}]/u_\tau$ for the same method and different Reynolds numbers. }
\label{fig:DU}
\end{figure}

\Fref{fig:DU} shows the $D[\tilde{U}]$ found by the UMZ-TNTI method and the local TKE method for three different stations studied. The figure shows that the $D[\tilde{U}]$ scales well with $u_\tau$ for both methods across the Reynolds numbers studied. In addition, the jump in velocity found by the UMZ-TNTI method is about 50\% larger than that found by the local TKE method, indicating that this property of the TNTI is more prominent in the interface found by the UMZ-TNTI method.

Following the analysis by \textcite{Chauhan2014b}, the momentum balance within the TNTI in the streamwise direction can be expressed using the thin shear layer equation:

\begin{equation}
U \frac{\partial U}{\partial x} + V \frac{\partial U}{\partial y} = -\frac{1}{\rho} \frac{\partial P}{\partial x} + \nu \frac{\partial^2 U}{\partial y^2} - \frac{\partial \overline{u^{\prime} v^{\prime}}}{\partial y}.
\label{equ:thin_shear_layer}
\end{equation}

\noindent In this equation, it is assumed that the majority of the TNTI is parallel to the $x$ direction, therefore $U$ dominates the tangential velocity along the interface and $V$ dominates the normal velocity across the interface. Assuming homogeneity in the tangential direction, $\partial / \partial x \approx 0$, and neglecting the turbulent fluctuation term, the $V \frac{\partial U}{\partial y}$ and $\nu \frac{\partial^2 U}{\partial y^2}$ terms in \eref{equ:thin_shear_layer} must balance. \Fref{fig:DU} shows that $\Delta U = \mathcal{O}(u_\tau)$ and \Fref{fig:mean_v} shows that $V = \mathcal{O}(0.1u_\tau)$, therefore the momentum balance suggests that the relevant length scale within the TNTI needs to satisfy the following order of magnitude relationship:

\begin{equation}
\mathcal{O}\left(0.1u_\tau \frac{u_\tau}{l} \right) = \mathcal{O}\left(\nu \frac{u_\tau}{l^2} \right).
\end{equation}

\noindent This equation suggests that the relevant length scale within the TNTI needs to be of the order of $\mathcal{O}(10l^+)$, where $l^+ = u_\tau / \nu$ is the viscous length scale of the flow. Therefore, $\frac{d \tilde{U}}{d \tilde{y}}|_{\mbox{\small{max}}}$ scales with $u_\tau / l^+$ and $\frac{\delta}{u_\tau} \frac{d \tilde{U}}{d \tilde{y}}|_{\mbox{\small{max}}}$ scales with $Re_\tau$.

\begin{figure}[!htbp]
\includegraphics[width=3.37 in]{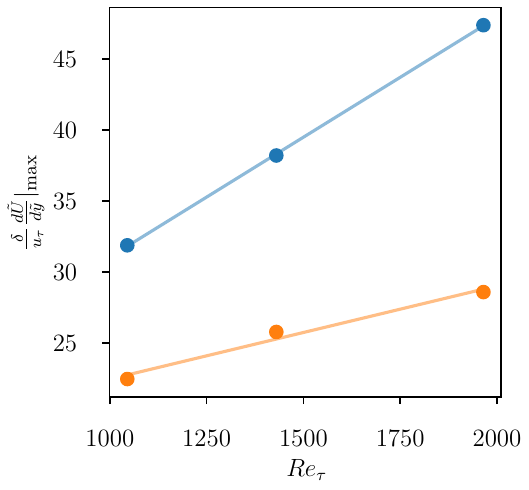} 
\caption{ The maximum velocity gradient within the TNTI, $\frac{d \tilde{U}}{d\tilde{y}}|_{\mbox{\small{max}}}$, normalised by $\frac{u_\tau}{\delta}$. See \Fref{fig:DU} for legends. Faint coloured lines represent the linear fit of the data points for the same method and different Reynolds numbers. }
\label{fig:dUdy_max}
\end{figure}

This relationship holds irrespective of the TNTI identification method used, and it is examined in \Fref{fig:dUdy_max} for the UMZ-TNTI method and the local TKE method. The figure shows that $\frac{\delta}{u_\tau} \frac{d \tilde{U}}{d \tilde{y}}|_{\mbox{\small{max}}}$ indeed has a linear relationship with $Re_\tau$ for both TNTI identification methods across the Reynolds number range studied. The figure also shows that the normalised maximum gradient found by the UMZ-TNTI method is about 50\% higher than that found using the local TKE method, indicating that the flow undergoes a sharper change within the interface.

\begin{figure}[!htbp]
\includegraphics[width=3.37 in]{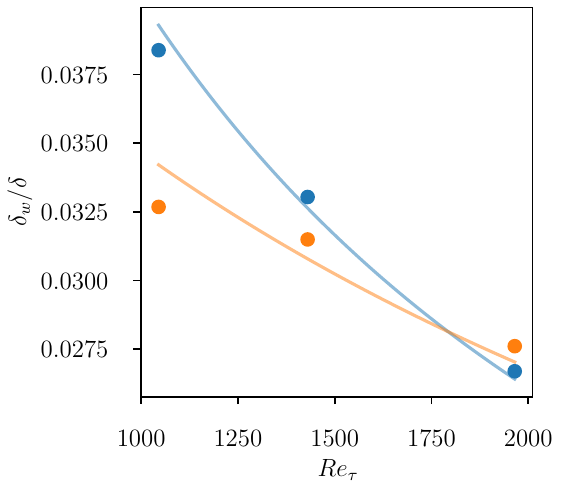} 
\caption{ The maximum velocity gradient within the TNTI, $\frac{d \tilde{U}}{d\tilde{y}}|_{\mbox{\small{max}}}$, normalised by $\frac{u_\tau}{\delta}$. See \Fref{fig:DU} for legends. Faint coloured lines represent \eref{equ:delta_omega_relationship}. }
\label{fig:thickness}
\end{figure}

Furthermore, the vorticity thickness, $\delta_\omega$, of the interface normalised by the boundary layer thickness is presented in \Fref{fig:thickness}. From the definition of $\delta_\omega$, the relationship between $\delta_\omega/\delta$ and $Re_\tau$ can be expressed as

\begin{equation}
\frac{\delta_\omega}{\delta} = \frac{D[\tilde{U}]/u_\tau}{a_1 Re_\tau + a_2},
\label{equ:delta_omega_relationship}
\end{equation}

\noindent where $a_1$ and $a_2$ are the parameters of the linear fit in \Fref{fig:dUdy_max}. This relationship is also plotted in \Fref{fig:thickness} and shows good agreement with the data points, as expected. The figure shows that the interface thickness grows slower than the boundary layer thickness, so that the ratio decreases as the Reynolds number increases.

The analysis of the conditionally averaged mean streamwise velocity shows that the TNTI found by the UMZ-TNTI method and the local TKE method have the same thickness, but as the velocity difference is 50\% higher in the TNTI found by the UMZ-TNTI method, the maximum velocity gradient is also 50\% higher. This higher velocity jump fits better with the conceptual model of the TNTI, where the mean streamwise velocity exhibits a rapid increase across the interface over a small distance. This process also identifies $D[\tilde{U}]$ as the appropriate velocity scale, as well as $\delta_\omega$ as the appropriate length scale for the flow within the TNTI. However, as a jump is not observed for the $\tilde{U}$ profiles resulting from the vorticity threshold method, these velocity and length scales do not apply to the TNTI identified using the vorticity threshold method.

\begin{figure}[!htbp]
\captionsetup[subfigure]{aboveskip=-1pt,belowskip=-1pt}
\centering
\begin{tabular}{cc}
\begin{subfigure}{0.4\textwidth}
\includegraphics[width=\textwidth]{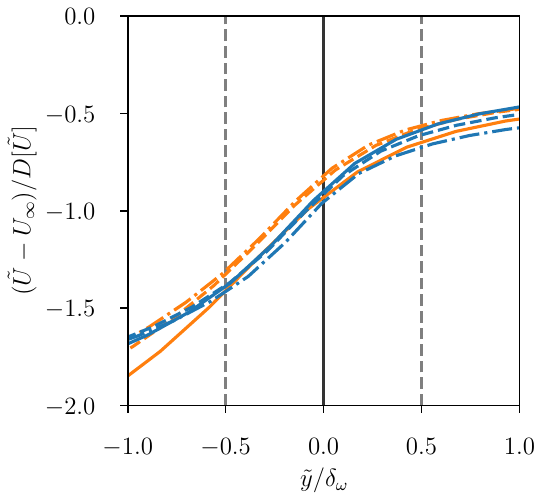} 
\caption{}
\label{fig:mean_u_DU}
\end{subfigure}

\begin{subfigure}{0.4\textwidth}
\includegraphics[width=\textwidth]{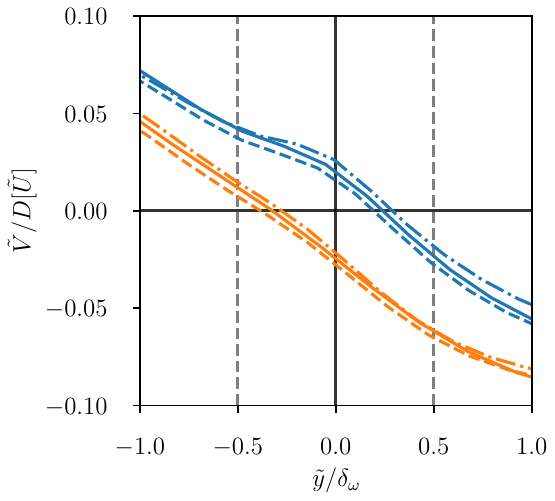} 
\caption{}
\label{fig:mean_v_DU}
\end{subfigure}
\end{tabular}
\vspace{-1.5em}
\caption{ Conditionally averaged mean velocities with respect to the \ac{TNTI}, scaled by velocity and length scales of the TNTI (a) Streamwise velocity, $(\tilde{U}-U_\infty)/D[\tilde{U}]$ (b) Wall normal velocity, $\tilde{V}/D[\tilde{U}]$. \textcolor{plt1}{\line}  UMZ-TNTI method, \textcolor{plt2}{\line} Local TKE method using 2C definition of $\tilde{k}$, \dotdash Station 1, \dashed Station 2, \line Station 3; Gray dashed lines mark the boundaries of the TNTI ($\pm 0.5 \delta_\omega$). }
\label{fig:cond_stat_mean_DU}
\end{figure}

The $\tilde{U}$ and $\tilde{V}$ profiles are plotted again using $D[\tilde{U}]$ and $\delta_\omega$ scaling in \Fref{fig:cond_stat_mean_DU}. The $\tilde{U}$ profiles in \Fref{fig:mean_u_DU} show good collapse between the UMZ-TNTI method and the local TKE method across the entire Reynolds number range. The collapse is observed not only within the interface but also above the interface, up to 2 interface thicknesses. The $\tilde{V}$ profiles in \Fref{fig:mean_v_DU} collapse for different Reynolds numbers for each method. However, there is a small offset of $0.05D[\tilde{U}]$ between the collapsed profiles found by the UMZ-TNTI method and the local TKE method.

\FloatBarrier
\subsection{Conditionally averaged Reynolds stress profiles}
In this section, the conditionally averaged Reynolds stress profiles are presented and discussed. As shown in the previous section, $D[\tilde{U}]$ is used as a velocity scale for the flow around the TNTI, so the Reynolds stress profiles are normalised by $D[\tilde{U}]^2$. Additionally, because there is no velocity jump across the interface identified using the vorticity threshold method, $D[\tilde{U}]$ and $\delta_\omega$ do not apply. Therefore, the Reynolds stress profiles from the vorticity threshold method are normalised by mean free stream velocity $U_\infty$ and $\delta$, which only allows qualitative comparison with the UMZ-TNTI method or the local TKE method.

%%%Figure moved so it is closer to the discussion
\begin{figure}[!htbp]
\captionsetup[subfigure]{aboveskip=-1pt,belowskip=-1pt}
\centering
\begin{tabular}{cc}
\begin{subfigure}{0.49\textwidth}
\includegraphics[width=\textwidth]{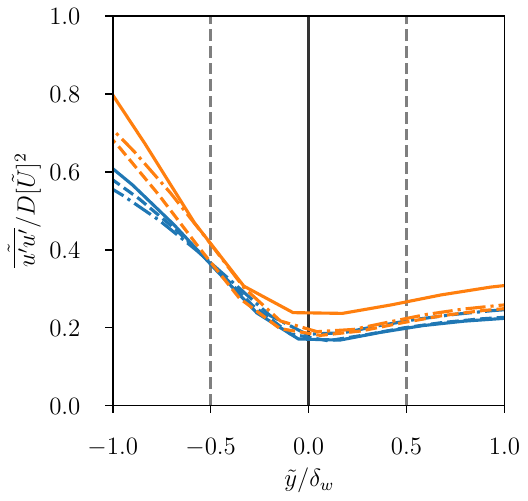} 
\caption{}
\label{fig:uu_UMZ_TKE}
\end{subfigure}
\begin{subfigure}{0.49\textwidth}
\includegraphics[width=\textwidth]{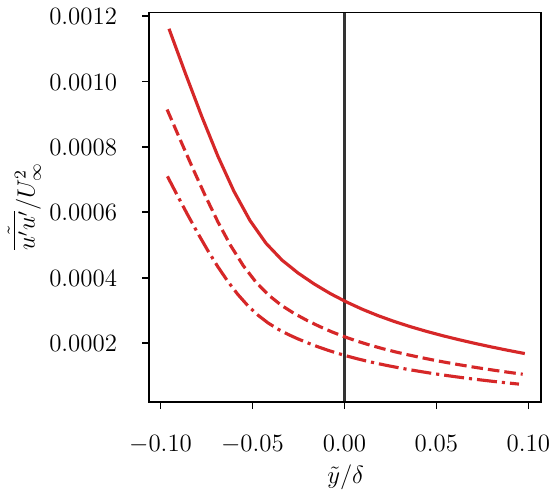} 
\caption{}
\label{fig:uu_vort}
\end{subfigure}
\end{tabular}
\vspace{-1.5em}
\caption{ Conditionally averaged Reynolds streamwise stress profiles, $\tilde{\overline{u'u'}}$. (a) Comparison of profiles obtained using the local TKE method and the UMZ–TNTI method. (b) Profile computed using the vorticity threshold method. \textcolor{plt1}{\line}  UMZ-TNTI method, \textcolor{plt2}{\line} Local TKE method using 2C definition of $\tilde{k}$, \textcolor{plt4}{\line} Vorticity threshold method, \dotdash Station 1, \dashed Station 2, \line Station 3.}
\label{fig:uu}
\end{figure}

The conditionally averaged Reynolds streamwise stress profiles, $\tilde{\overline{u'u'}}$, are presented in \Fref{fig:uu}. \Fref{fig:uu_UMZ_TKE} compares the profiles from the local TKE method and the UMZ-TNTI method, showing good agreement across the Reynolds number range. Above the TNTI, the normalised streamwise velocity fluctuation,$\tilde{\overline{u'u'}}/D[\tilde{U}]^2$ is significantly lower than below the TNTI, consistent with the definition of the TNTI, which separates the turbulent region with high-velocity fluctuation from the non-turbulent region. At $\tilde{y}=0$, the profile value represents the velocity fluctuation along the interface, which for the UMZ-TNTI method equals $\sigma_{u_{edge}}^2$. The figure shows that although different methods are used to identify the TNTI, the fluctuation along the interface scales well with $D[\tilde{U}]^2$ and is smaller than in the non-turbulent region. \Fref{fig:uu_vort} shows the profile from the vorticity threshold method, where the fluctuation above the interface is smaller than below, giving the profile an elbow-like shape and no local minimum at the interface.

\begin{figure}[!htbp]
\captionsetup[subfigure]{aboveskip=-1pt,belowskip=-1pt}
\centering
\begin{tabular}{cc}
\begin{subfigure}{0.49\textwidth}
\includegraphics[width=\textwidth]{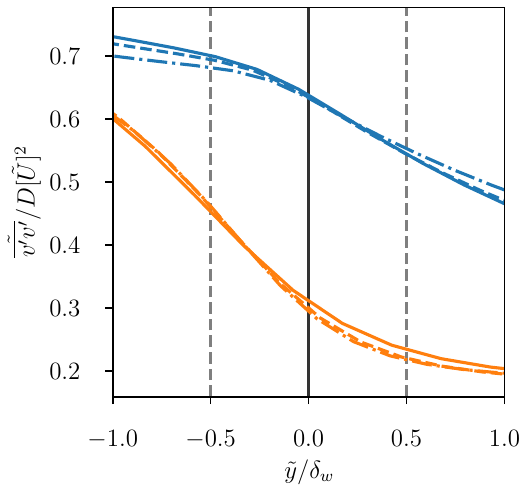} 
\caption{}
\label{fig:vv_UMZ_TKE}
\end{subfigure}
\begin{subfigure}{0.49\textwidth}
\includegraphics[width=\textwidth]{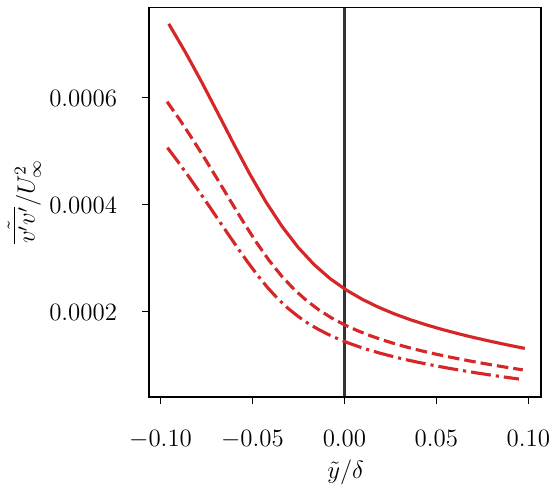} 
\caption{ }
\label{fig:vv_vort}
\end{subfigure}

\end{tabular}
\vspace{-1.5em}
\caption{ Conditionally averaged Reynolds wall normal stress profiles, $\tilde{\overline{v'v'}}$ (a) Comparison of profiles obtained using the local TKE method and the UMZ–TNTI method. (b) Profile computed using the vorticity threshold method. Legends are consistent with those in \Fref{fig:uu}.}
\label{fig:vv}
\end{figure}

The conditionally averaged Reynolds wall-normal stress profiles, $\tilde{\overline{v'v'}}$, are presented in \Fref{fig:vv}. \Fref{fig:vv_UMZ_TKE} shows that both local-TKE and UMZ-TNTI method collapse the $\tilde{\overline{v'v'}}$ profiles across the Reynolds number studied, however the collapsed profile from TNTI-UMZ method has a larger value than that from the local TKE method. The figure also shows that the change in the $\tilde{\overline{u'u'}}$ profile is sharper across the interface than for $\tilde{\overline{v'v'}}$, indicating different effects of the TNTI on streamwise and wall-normal fluctuations.

\begin{figure}[!htbp]
\captionsetup[subfigure]{aboveskip=-1pt,belowskip=-1pt}
\centering
\begin{tabular}{cc}
\begin{subfigure}{0.49\textwidth}
\includegraphics[width=\textwidth]{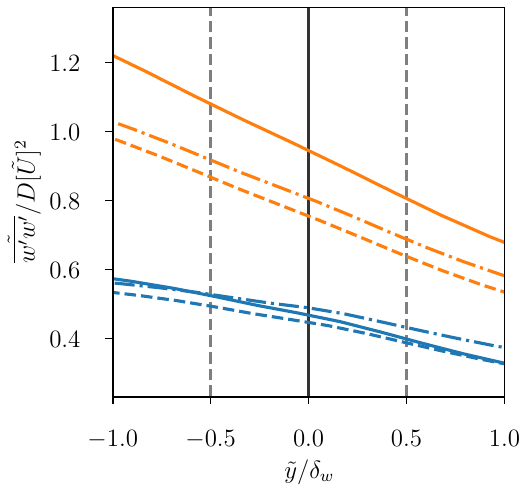} 
\caption{}
\label{fig:ww_UMZ_TKE}
\end{subfigure}
\begin{subfigure}{0.49\textwidth}
\includegraphics[width=\textwidth]{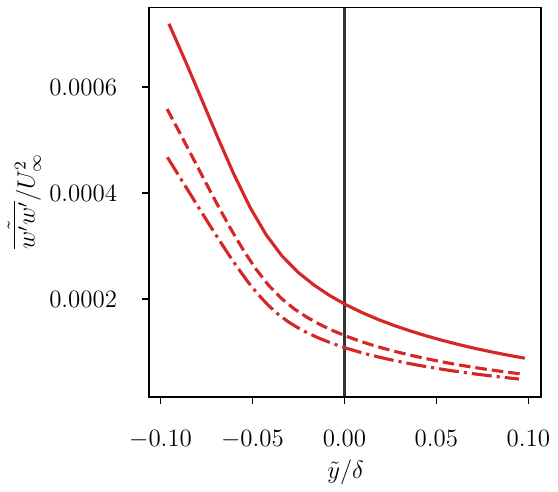} 
\caption{}
\label{fig:ww_vort}
\end{subfigure}
\end{tabular}
\vspace{-1.5em}
\caption{ Conditionally averaged Reynolds spanwise stress profiles, $\tilde{\overline{w'w'}}$ (a) Comparison of profiles obtained using the local TKE method and the UMZ–TNTI method. (b) Profile computed using the vorticity threshold method. Legends are consistent with those in \Fref{fig:uu}.}
\label{fig:ww}
\end{figure}

The conditionally averaged Reynolds spanwise stress profiles, $\tilde{\overline{w'w'}}$, are presented in \Fref{fig:ww}. \Fref{fig:ww_UMZ_TKE} shows that $\tilde{\overline{w'w'}}$ follows a similar trend above, within, and below the TNTI, without sharp changes. This reinforces the idea that the TNTI separates the turbulent and non-turbulent regions with varying effect for the three directions of Reynolds normal stresses: the streamwise fluctuation experiences the greatest reduction, wall-normal fluctuation shows moderate reduction, and spanwise fluctuation shows minimal reduction. This also explains why the 2C and 3C definitions of TKE produce nearly identical TNTIs. The difference between the 2C and 3C definitions lies in the spanwise velocity fluctuation term, which is less affected by the TNTI and, therefore, does not contribute to differentiating the turbulent and non-turbulent parts of the flow. Additionally, the spanwise stress profiles fail to scale with $D[\tilde{U}]^2$ for the local TKE method, but the profiles for different Reynolds number collapses better with UMZ-TNTI method, indicating that the UMZ-TNTI method can better capture the dynamics of velocity fluctuation in all three dimensions.

In contrast to the profiles from the local TKE and UMZ-TNTI methods, the Reynolds normal stress profiles from the vorticity threshold method, shown in \Fref{fig:uu_vort}, \Fref{fig:vv_vort}, and \Fref{fig:ww_vort}, are very similar in shape and amplitude. This suggests that the TNTI identified using the vorticity threshold method affects all three components similarly. This could be partly due to the TNTI being further from the wall compared to other methods, where turbulence is more isotropic.

%% Also, this suggests that the vorticity criterion exhibits less bias towards a particular direction in the velocity fluctuation.??

\begin{figure}[!htbp]
\captionsetup[subfigure]{aboveskip=-1pt,belowskip=-1pt}
\centering
\begin{tabular}{cc}
\begin{subfigure}{0.49\textwidth}
\includegraphics[width=\textwidth]{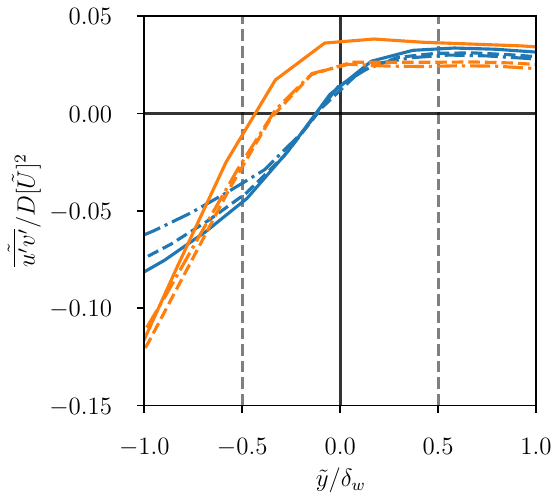} 
\caption{}
\label{fig:uv_UMZ_TKE}
\end{subfigure}
\begin{subfigure}{0.49\textwidth}
\includegraphics[width=\textwidth]{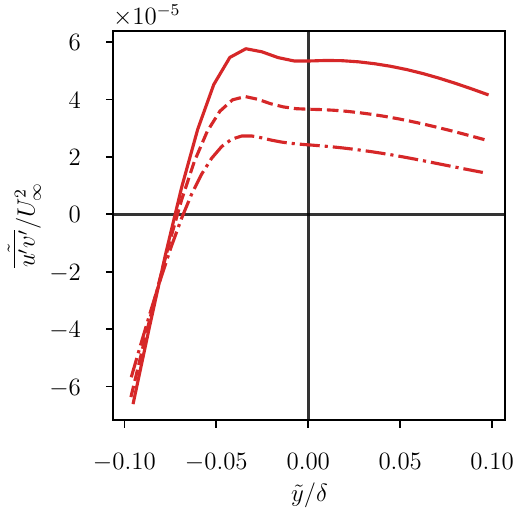} 
\caption{}
\label{fig:uv_vort}
\end{subfigure}
\end{tabular}
\vspace{-1.5em}
\caption{ Conditionally averaged Reynolds shear stress profiles, $\tilde{\overline{u'v'}}$ (a) Comparison of profiles obtained using the local TKE method and the UMZ–TNTI method. (b) Profile computed using the vorticity threshold method. Legends are consistent with those in \Fref{fig:uu}.}
\label{fig:uv}
\end{figure}

The conditionally averaged Reynolds shear stress profiles, $\tilde{\overline{u'v'}}$, are presented in \Fref{fig:uv}. \Fref{fig:uv_UMZ_TKE} shows a significant difference in the magnitude of $\tilde{\overline{u'v'}}$ above and below the TNTI, consistent with the $\tilde{\overline{u'u'}}$ profile and the definition of TNTI. The $\tilde{\overline{u'v'}}$ also undergoes a sign change, transitioning from negative in the turbulent region to positive in the non-turbulent region. In the case of the TNTI identified by the vorticity threshold method, $\tilde{\overline{u'v'}}$ undergoes a similar magnitude and sign change across the interface, though the position where the sign change occurs is below the TNTI, within the turbulent region.

\FloatBarrier
\subsection{Conditioned mean and fluctuation vorticity profiles}

In this section, the conditionally averaged mean and fluctuation vorticity profiles are presented. As mentioned previously, the TNTI-related length and vorticity scales are based on shear flow-like behaviour, which does not apply to the TNTI identified by the vorticity threshold method. Therefore, a quantitative comparison with the results obtained by the UMZ-TNTI method and the local TKE method is not possible. In this section, the vorticity profiles are normalised by star units as defined in \eref{equ:vorticity_star_notation}. 

\begin{figure}[!htbp]
\captionsetup[subfigure]{aboveskip=-1pt,belowskip=-1pt}
\centering
\begin{tabular}{cc}
\begin{subfigure}{0.49\textwidth}
\includegraphics[width=\textwidth]{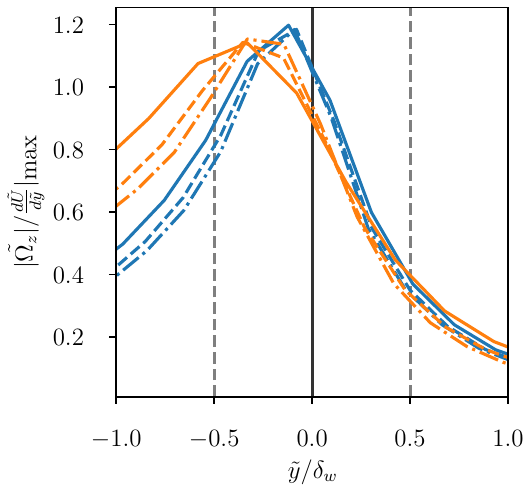} 
\caption{}
\label{fig:vortz_m_UMZ_TKE}
\end{subfigure}
\begin{subfigure}{0.49\textwidth}
\includegraphics[width=\textwidth]{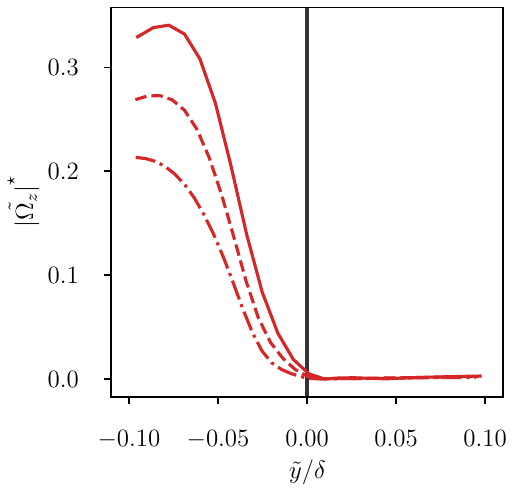} 
\caption{}
\label{fig:vortz_m_vort}
\end{subfigure}
\end{tabular}
\vspace{-1.5em}
\caption{ Conditionally averaged mean spanwise vorticity profiles, $|\tilde{\Omega_z}|$ (a) Comparison of profiles obtained using the local TKE method and the UMZ–TNTI method. (b) Profile computed using the vorticity threshold method. Legends are consistent with those in \Fref{fig:uu}.}
\label{fig:vortz_m}
\end{figure}

Conditionally averaged mean spanwise vorticity profiles, $|\tilde{\Omega_z}|$, are shown in \Fref{fig:vortz_m}. For the profiles obtained from both the local TKE method and the UMZ-TNTI method, the spanwise vorticity is concentrated within the interface thickness, while the UMZ-TNTI methods shows better collapse for different Reynolds number than the local TKE method. Also, the peak value of the normalised vorticity profiles is about 1.1, showing that majority of the mean spanwise vorticity is due to the steep gradient of $\tilde{U}$ profiles across the interface, which also verifies the assumption that the velocity field within the TNTI changes slowly with respect to the streamwise direction, leading to a small streamwise gradient. In contrast, the $|\tilde{\Omega_z}|$ profiles from the vorticity threshold method show a different behaviour, where the vorticity is concentrated below the TNTI rather than within it. This is expected, as the imposed threshold separates the turbulent region, characterised by high vorticity, from the low vorticity region.

The vorticity fluctuation profiles are presented in \Fref{fig:w_fluc}. The figure shows that the vorticity fluctuation profiles also collapse better across different Reynolds number for the UMZ-TNTI method than local TKE method. In addition, the TNTI alter the vorticity fluctuation profiles differently for different directions. The streamwise vorticity fluctuation, $\tilde{\overline{\omega_x'\omega_x'}}$, shows a sharp decrease within the interface, while the vorticity fluctuations in the spanwise and wall-normal directions, $\tilde{\overline{\omega_y'\omega_y'}}$ and $\tilde{\overline{\omega_z'\omega_z'}}$, exhibit a localised peak just below the interface in the turbulent region. The effect of the TNTI on the vorticity fluctuation profiles is more pronounced in the profiles produced by the UMZ-TNTI method, suggesting that the UMZ-TNTI method captures this effect better than the local TKE method. The conditioned vorticity fluctuation profiles obtained by the vorticity thresholding method, shown in \Fref{fig:w_fluc}(b, d, f), are very similar across the three directions, likely due to the more isotropic nature of the flow in this region. A rapid decrease in the magnitude of the vorticity fluctuation across the TNTI can also be observed.

%%Julio: Does this mean the streamwise vorticity fluctuations are transferred (re-oriented) to the spanwise and wall-normal directions?

\begin{figure}[!htbp]
\captionsetup[subfigure]{aboveskip=-1pt,belowskip=-1pt}
\centering
\begin{tabular}{cc}

\begin{subfigure}{0.49\textwidth}
\includegraphics[width=\textwidth]{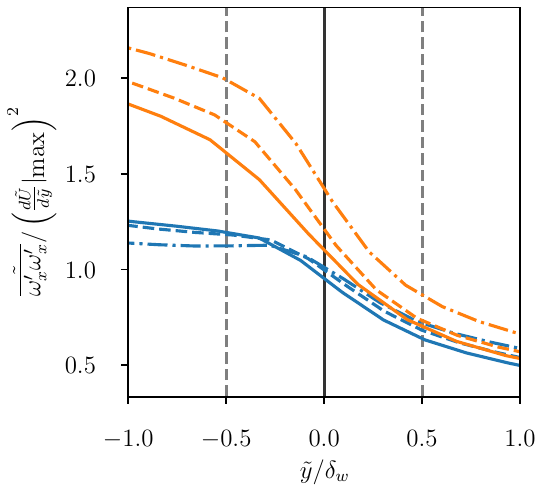} 
\caption{}
\label{fig:wxwx_UMZ_TKE}
\end{subfigure}
\begin{subfigure}{0.49\textwidth}
\includegraphics[width=\textwidth]{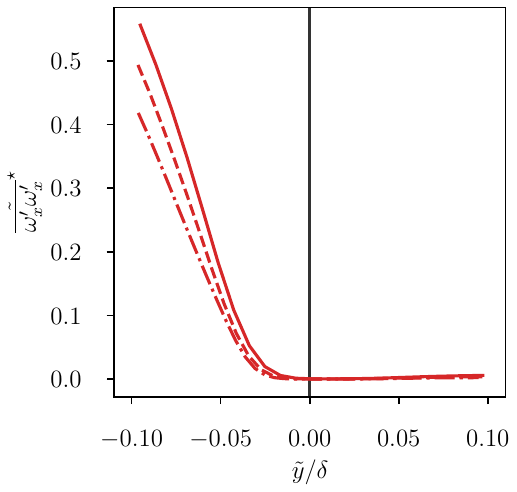} 
\caption{}
\label{fig:wxwx_vort}
\end{subfigure}

\end{tabular}
\vspace{-1.5em}

\caption{Conditioned averaged vorticity fluctuation profiles: (a, b) $\tilde{\overline{\omega_x'\omega_x'}}$, (c, d) $\tilde{\overline{\omega_y'\omega_y'}}$, (e, f) $\tilde{\overline{\omega_z'\omega_z'}}$. Panels (a, c, e) compare the profiles obtained by the UMZ-TNTI method and the local TKE method, normalised by $\frac{d\tilde{U}}{d\tilde{y}}|_{\mbox{max}}$. Panels (b, d, f) show the profiles obtained by the vorticity threshold method, normalised in star units. Legends are the same as in \Fref{fig:uu}.}

\end{figure}

\begin{figure}[!htbp]
\captionsetup[subfigure]{aboveskip=-1pt,belowskip=-1pt}
\ContinuedFloat
\centering
\begin{tabular}{cc}

\begin{subfigure}{0.49\textwidth}
\includegraphics[width=\textwidth]{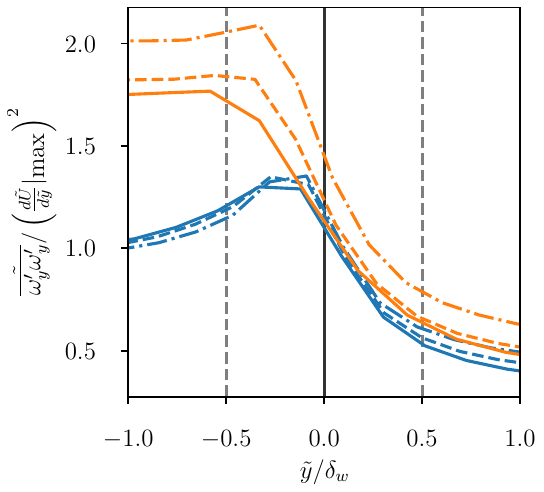} 
\caption{}
\label{fig:wywy_UMZ_TKE}
\end{subfigure}
\begin{subfigure}{0.49\textwidth}
\includegraphics[width=\textwidth]{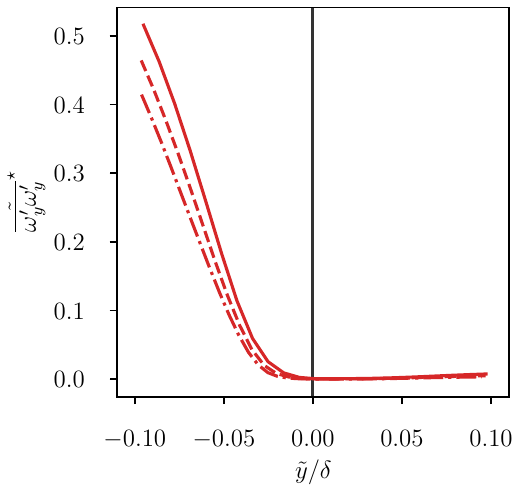} 
\caption{}
\label{fig:wywy_vort}
\end{subfigure}

\\

\begin{subfigure}{0.49\textwidth}
\includegraphics[width=\textwidth]{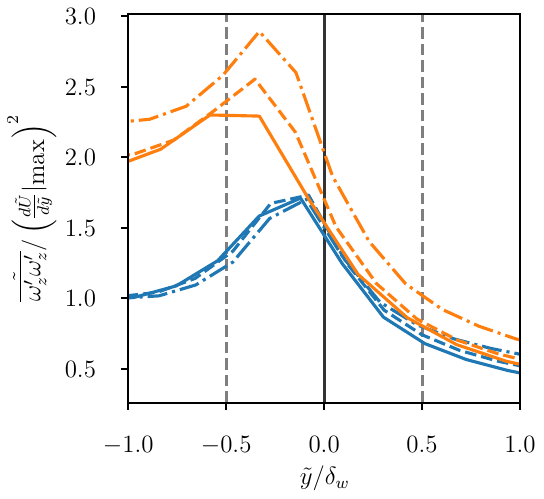} 
\caption{}
\label{fig:wzwz_UMZ_TKE}
\end{subfigure}
\begin{subfigure}{0.49\textwidth}
\includegraphics[width=\textwidth]{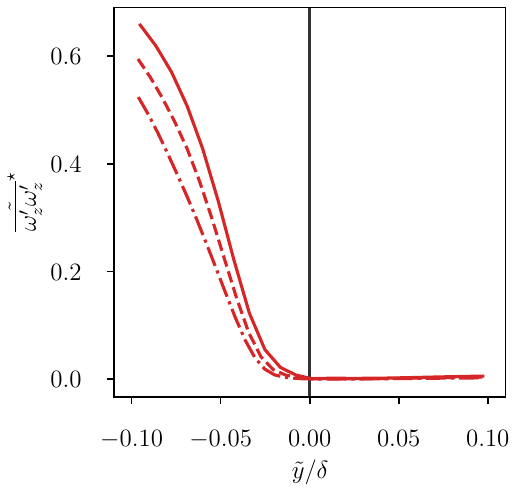} 
\caption{}
\label{fig:wzwz_vort}
\end{subfigure}

\end{tabular}
\vspace{-1.5em}
\caption{ Conditionally averaged vorticity fluctuation profiles (a,b) $\tilde{\overline{\omega_x'\omega_x'}}$ (c,d) $\tilde{\overline{\omega_y'\omega_y'}}$ (e,f) $\tilde{\overline{\omega_z'\omega_z'}}$. (a,c,e) compares the profiles acquired by the UMZ-TNTI method and the local TKE method, which are normalised by $\frac{d\tilde{U}}{d\tilde{y}}|_{\mbox{max}}$. (b,d,f) are the profiles acquired by the vorticity threshold method, and are in star unit. Legends of the figures are the same as \Fref{fig:uu}.}
\label{fig:w_fluc}
\end{figure}

\FloatBarrier
\section{Conclusion}

In this paper, a novel methodology for identifying the turbulent/non-turbulent interface (TNTI) in turbulent flows is developed using the UMZ concept. The streamwise velocity PDF is calculated within a sliding window, where $u_{edge}$ separating the turbulent and non-turbulent part of the flow is identified as the local minima closest to the free stream.  This approach eliminates the need for manually selected thresholds, allowing for more objective and consistent identification of TNTIs across different experiments and simulations. 

The geometric properties of the identified TNTI are analysed, focusing on the probability density function of the interface height, the intermittency profile, and the mean and standard deviation of the interface height. The study reveals that the TNTI scales well with the local boundary layer thickness and shows only minor variations across the Reynolds number range from $Re_\tau = 1,000$ to $2,000$. A sensitivity analysis of the streamwise extent of the sliding window demonstrates that the domain length has minimal impact on the TNTI identified.

When compared to existing methods, such as the local TKE method and the vorticity threshold method, the UMZ-TNTI methodology stands out by not requiring a threshold, making it more universally applicable. The TNTI identified using the local TKE method partially overlaps with the TNTI identified by the UMZ-TNTI method, while the vorticity threshold method identifies a more different interface, located farther from the wall. Furthermore, the mean and standard deviation of the TNTI are consistent with other studies that do not rely on vorticity-based methods.

The conditional turbulent statistics based on the TNTI height show results. The mean streamwise velocity profile exhibits high shear within the TNTI, which is characteristic of mixing layer behaviour, and suggests velocity and length scales local to the TNTI. The velocity fluctuation profiles reveal varying effects across different directions, with $\tilde{\overline{u'u'}}$ showing large differences, $\tilde{\overline{v'v'}}$ showing smaller differences, and $\tilde{\overline{w'w'}}$ showing negligible effects. Both mean and fluctuating vorticity profiles show better collapse for the UMZ-TNTI method than the local TKE method. The mean spanwise vorticity concentrates within the TNTI identified by the UMZ-TNTI method, in contrast to the vorticity threshold method, where the vorticity is concentrated on the turbulent side of the flow. Streamwise vorticity fluctuation decreases across the interface, while vorticity fluctuations in the other directions exhibit localised peaks within the interface.

Although the local TKE method remains useful for identifying envelopes of elevated turbulence intensity, the UMZ-TNTI method more sharply isolates the interfacial region where the most significant velocity and vorticity gradients occur. This renders it a better diagnostic tool for entrainment and mixing-layer studies, particularly in high-resolution PIV where TKE estimates are limited by spatial resolution and noise.

% If in two-column mode, this environment will change to single-column format so that long equations can be displayed. 
%\begin{widetext}
%$$\mbox{put long equation here}$$
%\end{widetext}

% Figures should be put into the text as floats. 
% Use the graphics or graphicx packages (distributed with LaTeX2e).
% Put the label that you will use with \ref{} command in the braces of the \label{} command.
%
% \begin{figure}
% \end{figure}

% \begin{table}
% \end{tabular}
% \end{table}

\begin{acknowledgments}
The authors would like to acknowledge the funding of this research by the Australian Research Council through a Discovery Grant. The authors also acknowledge the computational resources provided by the Pawsey Supercomputing Centre supported by the Australian and Western Australian Governments, and the National Computational Infrastructure (NCI), supported by the Australian Government through computational grants awarded by the National Computational Merit Allocation Scheme (NCMAS) funded by the Australian Government, as well as a Monash HPC-NCI Merit Allocation. Bihai Sun gratefully acknowledges the support provided by a Monash Graduate Scholarship (MGS).\\
The authors also would like to acknowledge the Fluid Dynamics Group in Universidad Polit\`{e}cnica de Madrid, led by Prof. Javier Jim\'{e}nez, for providing the DNS dataset used in this study.
\end{acknowledgments}

\bibliography{TNTI_UMZ}

\begin{thebibliography}{33}%
\makeatletter
\providecommand \@ifxundefined [1]{%
 \@ifx{#1\undefined}
}%
\providecommand \@ifnum [1]{%
 \ifnum #1\expandafter \@firstoftwo
 \else \expandafter \@secondoftwo
 \fi
}%
\providecommand \@ifx [1]{%
 \ifx #1\expandafter \@firstoftwo
 \else \expandafter \@secondoftwo
 \fi
}%
\providecommand \natexlab [1]{#1}%
\providecommand \enquote  [1]{``#1''}%
\providecommand \bibnamefont  [1]{#1}%
\providecommand \bibfnamefont [1]{#1}%
\providecommand \citenamefont [1]{#1}%
\providecommand \href@noop [0]{\@secondoftwo}%
\providecommand \href [0]{\begingroup \@sanitize@url \@href}%
\providecommand \@href[1]{\@@startlink{#1}\@@href}%
\providecommand \@@href[1]{\endgroup#1\@@endlink}%
\providecommand \@sanitize@url [0]{\catcode `\\12\catcode `\$12\catcode
  `\&12\catcode `\#12\catcode `\^12\catcode `\_12\catcode `\%12\relax}%
\providecommand \@@startlink[1]{}%
\providecommand \@@endlink[0]{}%
\providecommand \url  [0]{\begingroup\@sanitize@url \@url }%
\providecommand \@url [1]{\endgroup\@href {#1}{\urlprefix }}%
\providecommand \urlprefix  [0]{URL }%
\providecommand \Eprint [0]{\href }%
\providecommand \doibase [0]{https://doi.org/}%
\providecommand \selectlanguage [0]{\@gobble}%
\providecommand \bibinfo  [0]{\@secondoftwo}%
\providecommand \bibfield  [0]{\@secondoftwo}%
\providecommand \translation [1]{[#1]}%
\providecommand \BibitemOpen [0]{}%
\providecommand \bibitemStop [0]{}%
\providecommand \bibitemNoStop [0]{.\EOS\space}%
\providecommand \EOS [0]{\spacefactor3000\relax}%
\providecommand \BibitemShut  [1]{\csname bibitem#1\endcsname}%
\let\auto@bib@innerbib\@empty
%</preamble>
\bibitem [{\citenamefont {da~Silva}\ \emph {et~al.}(2014)\citenamefont
  {da~Silva}, \citenamefont {Hunt}, \citenamefont {Eames},\ and\ \citenamefont
  {Westerweel}}]{Carlos2014}%
  \BibitemOpen
  \bibfield  {author} {\bibinfo {author} {\bibfnamefont {C.~B.}\ \bibnamefont
  {da~Silva}}, \bibinfo {author} {\bibfnamefont {J.~C.}\ \bibnamefont {Hunt}},
  \bibinfo {author} {\bibfnamefont {I.}~\bibnamefont {Eames}},\ and\ \bibinfo
  {author} {\bibfnamefont {J.}~\bibnamefont {Westerweel}},\ }\href
  {https://doi.org/10.1146/annurev-fluid-010313-141357} {\bibfield  {journal}
  {\bibinfo  {journal} {Annual Review of Fluid Mechanics}\ }\textbf {\bibinfo
  {volume} {46}},\ \bibinfo {pages} {567} (\bibinfo {year} {2014})}\BibitemShut
  {NoStop}%
\bibitem [{\citenamefont {Corrsin}\ and\ \citenamefont
  {Kistler}(1955)}]{Corrsin1955}%
  \BibitemOpen
  \bibfield  {author} {\bibinfo {author} {\bibfnamefont {S.}~\bibnamefont
  {Corrsin}}\ and\ \bibinfo {author} {\bibfnamefont {A.~L.}\ \bibnamefont
  {Kistler}},\ }\href@noop {} {\emph {\bibinfo {title} {Free-Stream Boundaries
  of Turbulent Flows}}},\ \bibinfo {type} {Tech. Rep.}\ \bibinfo {number}
  {NACA-TN-3133}\ (\bibinfo  {institution} {The Johns Hopkins University},\
  \bibinfo {year} {1955})\BibitemShut {NoStop}%
\bibitem [{\citenamefont {Bisset}\ \emph {et~al.}(2002)\citenamefont {Bisset},
  \citenamefont {Hunt},\ and\ \citenamefont {Rogers}}]{Bisset2002}%
  \BibitemOpen
  \bibfield  {author} {\bibinfo {author} {\bibfnamefont {D.~K.}\ \bibnamefont
  {Bisset}}, \bibinfo {author} {\bibfnamefont {J.~C.~R.}\ \bibnamefont
  {Hunt}},\ and\ \bibinfo {author} {\bibfnamefont {M.~M.}\ \bibnamefont
  {Rogers}},\ }\href {https://doi.org/10.1017/S0022112001006759} {\bibfield
  {journal} {\bibinfo  {journal} {Journal of Fluid Mechanics}\ }\textbf
  {\bibinfo {volume} {451}},\ \bibinfo {pages} {383} (\bibinfo {year}
  {2002})}\BibitemShut {NoStop}%
\bibitem [{\citenamefont {Westerweel}\ \emph {et~al.}(2009)\citenamefont
  {Westerweel}, \citenamefont {Fukushima}, \citenamefont {Pedersen},\ and\
  \citenamefont {Hunt}}]{Westerweel2009}%
  \BibitemOpen
  \bibfield  {author} {\bibinfo {author} {\bibfnamefont {J.}~\bibnamefont
  {Westerweel}}, \bibinfo {author} {\bibfnamefont {C.}~\bibnamefont
  {Fukushima}}, \bibinfo {author} {\bibfnamefont {J.~M.}\ \bibnamefont
  {Pedersen}},\ and\ \bibinfo {author} {\bibfnamefont {J.~C.~R.}\ \bibnamefont
  {Hunt}},\ }\href {https://doi.org/10.1017/S0022112009006600} {\bibfield
  {journal} {\bibinfo  {journal} {Journal of Fluid Mechanics}\ }\textbf
  {\bibinfo {volume} {631}},\ \bibinfo {pages} {199} (\bibinfo {year}
  {2009})}\BibitemShut {NoStop}%
\bibitem [{\citenamefont {Holzner}\ and\ \citenamefont
  {Lüthi}(2011)}]{Holzner2011}%
  \BibitemOpen
  \bibfield  {author} {\bibinfo {author} {\bibfnamefont {M.}~\bibnamefont
  {Holzner}}\ and\ \bibinfo {author} {\bibfnamefont {B.}~\bibnamefont
  {Lüthi}},\ }\href {https://doi.org/10.1103/PhysRevLett.106.134503}
  {\bibfield  {journal} {\bibinfo  {journal} {Physical Review Letters}\
  }\textbf {\bibinfo {volume} {106}},\ \bibinfo {pages} {134503} (\bibinfo
  {year} {2011})}\BibitemShut {NoStop}%
\bibitem [{\citenamefont {Fiedler}\ and\ \citenamefont
  {Head}(1966)}]{Fiedler1966}%
  \BibitemOpen
  \bibfield  {author} {\bibinfo {author} {\bibfnamefont {H.}~\bibnamefont
  {Fiedler}}\ and\ \bibinfo {author} {\bibfnamefont {M.~R.}\ \bibnamefont
  {Head}},\ }\href {https://doi.org/10.1017/S0022112066000363} {\bibfield
  {journal} {\bibinfo  {journal} {Journal of Fluid Mechanics}\ }\textbf
  {\bibinfo {volume} {25}},\ \bibinfo {pages} {719} (\bibinfo {year}
  {1966})}\BibitemShut {NoStop}%
\bibitem [{\citenamefont {Townsend}(1980)}]{Townsend1980}%
  \BibitemOpen
  \bibfield  {author} {\bibinfo {author} {\bibfnamefont {A.~A.~R.}\
  \bibnamefont {Townsend}},\ }\href@noop {} {\emph {\bibinfo {title} {The
  Structure of Turbulent Shear Flow}}}\ (\bibinfo  {publisher} {Cambridge
  University Press},\ \bibinfo {year} {1980})\BibitemShut {NoStop}%
\bibitem [{\citenamefont {de~Silva}\ \emph {et~al.}(2013)\citenamefont
  {de~Silva}, \citenamefont {Philip}, \citenamefont {Chauhan}, \citenamefont
  {Meneveau},\ and\ \citenamefont {Marusic}}]{deSilva2013}%
  \BibitemOpen
  \bibfield  {author} {\bibinfo {author} {\bibfnamefont {C.~M.}\ \bibnamefont
  {de~Silva}}, \bibinfo {author} {\bibfnamefont {J.}~\bibnamefont {Philip}},
  \bibinfo {author} {\bibfnamefont {K.}~\bibnamefont {Chauhan}}, \bibinfo
  {author} {\bibfnamefont {C.}~\bibnamefont {Meneveau}},\ and\ \bibinfo
  {author} {\bibfnamefont {I.}~\bibnamefont {Marusic}},\ }\href
  {https://doi.org/10.1103/PhysRevLett.111.044501} {\bibfield  {journal}
  {\bibinfo  {journal} {Physical Review Letters}\ }\textbf {\bibinfo {volume}
  {111}},\ \bibinfo {pages} {044501} (\bibinfo {year} {2013})}\BibitemShut
  {NoStop}%
\bibitem [{\citenamefont {Borrell}\ and\ \citenamefont
  {Jiménez}(2016)}]{Borrell2016}%
  \BibitemOpen
  \bibfield  {author} {\bibinfo {author} {\bibfnamefont {G.}~\bibnamefont
  {Borrell}}\ and\ \bibinfo {author} {\bibfnamefont {J.}~\bibnamefont
  {Jiménez}},\ }\href {https://doi.org/10.1017/jfm.2016.430} {\bibfield
  {journal} {\bibinfo  {journal} {Journal of Fluid Mechanics}\ }\textbf
  {\bibinfo {volume} {801}},\ \bibinfo {pages} {554} (\bibinfo {year}
  {2016})}\BibitemShut {NoStop}%
\bibitem [{\citenamefont {Anand}\ \emph {et~al.}(2009)\citenamefont {Anand},
  \citenamefont {Boersma},\ and\ \citenamefont {Agrawal}}]{Anand2009}%
  \BibitemOpen
  \bibfield  {author} {\bibinfo {author} {\bibfnamefont {R.~K.}\ \bibnamefont
  {Anand}}, \bibinfo {author} {\bibfnamefont {B.~J.}\ \bibnamefont {Boersma}},\
  and\ \bibinfo {author} {\bibfnamefont {A.}~\bibnamefont {Agrawal}},\ }\href
  {https://doi.org/10.1007/s00348-009-0695-5} {\bibfield  {journal} {\bibinfo
  {journal} {Experiments in Fluids}\ }\textbf {\bibinfo {volume} {47}},\
  \bibinfo {pages} {995} (\bibinfo {year} {2009})}\BibitemShut {NoStop}%
\bibitem [{\citenamefont {Prasad}\ and\ \citenamefont
  {Sreenivasan}(1989)}]{Prasad1989}%
  \BibitemOpen
  \bibfield  {author} {\bibinfo {author} {\bibfnamefont {R.~R.}\ \bibnamefont
  {Prasad}}\ and\ \bibinfo {author} {\bibfnamefont {K.~R.}\ \bibnamefont
  {Sreenivasan}},\ }\href {https://doi.org/10.1007/BF00198005} {\bibfield
  {journal} {\bibinfo  {journal} {Experiments in Fluids}\ }\textbf {\bibinfo
  {volume} {7}},\ \bibinfo {pages} {259} (\bibinfo {year} {1989})}\BibitemShut
  {NoStop}%
\bibitem [{\citenamefont {da~Silva}\ \emph {et~al.}(2011)\citenamefont
  {da~Silva}, \citenamefont {dos Reis},\ and\ \citenamefont
  {Pereira}}]{deSilva2011}%
  \BibitemOpen
  \bibfield  {author} {\bibinfo {author} {\bibfnamefont {C.~B.}\ \bibnamefont
  {da~Silva}}, \bibinfo {author} {\bibfnamefont {R.~J.~N.}\ \bibnamefont {dos
  Reis}},\ and\ \bibinfo {author} {\bibfnamefont {J.~C.~F.}\ \bibnamefont
  {Pereira}},\ }\href {https://doi.org/10.1017/jfm.2011.296} {\bibfield
  {journal} {\bibinfo  {journal} {Journal of Fluid Mechanics}\ }\textbf
  {\bibinfo {volume} {685}},\ \bibinfo {pages} {165} (\bibinfo {year}
  {2011})}\BibitemShut {NoStop}%
\bibitem [{\citenamefont {Chauhan}\ \emph
  {et~al.}(2014{\natexlab{a}})\citenamefont {Chauhan}, \citenamefont {Philip},
  \citenamefont {de~Silva}, \citenamefont {Hutchins},\ and\ \citenamefont
  {Marusic}}]{Chauhan2014}%
  \BibitemOpen
  \bibfield  {author} {\bibinfo {author} {\bibfnamefont {K.}~\bibnamefont
  {Chauhan}}, \bibinfo {author} {\bibfnamefont {J.}~\bibnamefont {Philip}},
  \bibinfo {author} {\bibfnamefont {C.~M.}\ \bibnamefont {de~Silva}}, \bibinfo
  {author} {\bibfnamefont {N.}~\bibnamefont {Hutchins}},\ and\ \bibinfo
  {author} {\bibfnamefont {I.}~\bibnamefont {Marusic}},\ }\href
  {https://doi.org/10.1017/jfm.2013.641} {\bibfield  {journal} {\bibinfo
  {journal} {Journal of Fluid Mechanics}\ }\textbf {\bibinfo {volume} {742}},\
  \bibinfo {pages} {119} (\bibinfo {year} {2014}{\natexlab{a}})}\BibitemShut
  {NoStop}%
\bibitem [{\citenamefont {Meinhart}\ and\ \citenamefont
  {Adrian}(1995)}]{Meinhart1995}%
  \BibitemOpen
  \bibfield  {author} {\bibinfo {author} {\bibfnamefont {C.~D.}\ \bibnamefont
  {Meinhart}}\ and\ \bibinfo {author} {\bibfnamefont {R.~J.}\ \bibnamefont
  {Adrian}},\ }\href {https://doi.org/10.1063/1.868594} {\bibfield  {journal}
  {\bibinfo  {journal} {Physics of Fluids}\ }\textbf {\bibinfo {volume} {7}},\
  \bibinfo {pages} {694} (\bibinfo {year} {1995})}\BibitemShut {NoStop}%
\bibitem [{\citenamefont {Adrian}\ \emph {et~al.}(2000)\citenamefont {Adrian},
  \citenamefont {Meinhart},\ and\ \citenamefont {Tomkins}}]{Adrian2000}%
  \BibitemOpen
  \bibfield  {author} {\bibinfo {author} {\bibfnamefont {R.~J.}\ \bibnamefont
  {Adrian}}, \bibinfo {author} {\bibfnamefont {C.~D.}\ \bibnamefont
  {Meinhart}},\ and\ \bibinfo {author} {\bibfnamefont {C.~D.}\ \bibnamefont
  {Tomkins}},\ }\href {https://doi.org/10.1017/S0022112000001580} {\bibfield
  {journal} {\bibinfo  {journal} {Journal of Fluid Mechanics}\ }\textbf
  {\bibinfo {volume} {422}},\ \bibinfo {pages} {1} (\bibinfo {year}
  {2000})}\BibitemShut {NoStop}%
\bibitem [{\citenamefont {Laskari}\ \emph {et~al.}(2018)\citenamefont
  {Laskari}, \citenamefont {de~Kat}, \citenamefont {Hearst},\ and\
  \citenamefont {Ganapathisubramani}}]{Laskari2018}%
  \BibitemOpen
  \bibfield  {author} {\bibinfo {author} {\bibfnamefont {A.}~\bibnamefont
  {Laskari}}, \bibinfo {author} {\bibfnamefont {R.}~\bibnamefont {de~Kat}},
  \bibinfo {author} {\bibfnamefont {R.~J.}\ \bibnamefont {Hearst}},\ and\
  \bibinfo {author} {\bibfnamefont {B.}~\bibnamefont {Ganapathisubramani}},\
  }\href {https://doi.org/10.1017/jfm.2018.126} {\bibfield  {journal} {\bibinfo
   {journal} {Journal of Fluid Mechanics}\ }\textbf {\bibinfo {volume} {842}},\
  \bibinfo {pages} {554} (\bibinfo {year} {2018})}\BibitemShut {NoStop}%
\bibitem [{\citenamefont {Senthil}\ \emph {et~al.}(2020)\citenamefont
  {Senthil}, \citenamefont {Atkinson},\ and\ \citenamefont
  {Soria}}]{Senthil2020}%
  \BibitemOpen
  \bibfield  {author} {\bibinfo {author} {\bibfnamefont {S.}~\bibnamefont
  {Senthil}}, \bibinfo {author} {\bibfnamefont {C.}~\bibnamefont {Atkinson}},\
  and\ \bibinfo {author} {\bibfnamefont {J.}~\bibnamefont {Soria}},\ }\href
  {https://doi.org/10.1088/1742-6596/1522/1/012013} {\bibfield  {journal}
  {\bibinfo  {journal} {Journal of Physics: Conference Series}\ }\textbf
  {\bibinfo {volume} {1522}},\ \bibinfo {pages} {012013} (\bibinfo {year}
  {2020})}\BibitemShut {NoStop}%
\bibitem [{\citenamefont {de~Silva}\ \emph {et~al.}(2017)\citenamefont
  {de~Silva}, \citenamefont {Philip}, \citenamefont {Hutchins},\ and\
  \citenamefont {Marusic}}]{deSilva2017}%
  \BibitemOpen
  \bibfield  {author} {\bibinfo {author} {\bibfnamefont {C.~M.}\ \bibnamefont
  {de~Silva}}, \bibinfo {author} {\bibfnamefont {J.}~\bibnamefont {Philip}},
  \bibinfo {author} {\bibfnamefont {N.}~\bibnamefont {Hutchins}},\ and\
  \bibinfo {author} {\bibfnamefont {I.}~\bibnamefont {Marusic}},\ }\href
  {https://doi.org/10.1017/jfm.2017.197} {\bibfield  {journal} {\bibinfo
  {journal} {Journal of Fluid Mechanics}\ }\textbf {\bibinfo {volume} {820}},\
  \bibinfo {pages} {451} (\bibinfo {year} {2017})}\BibitemShut {NoStop}%
\bibitem [{\citenamefont {Sun}\ \emph {et~al.}(2024)\citenamefont {Sun},
  \citenamefont {Atkinson},\ and\ \citenamefont {Soria}}]{Sun2024}%
  \BibitemOpen
  \bibfield  {author} {\bibinfo {author} {\bibfnamefont {B.}~\bibnamefont
  {Sun}}, \bibinfo {author} {\bibfnamefont {C.}~\bibnamefont {Atkinson}},\ and\
  \bibinfo {author} {\bibfnamefont {J.}~\bibnamefont {Soria}},\ }in\ \href@noop
  {} {\emph {\bibinfo {booktitle} {1st European Fluid Dynamics Conference
  (EFDC1)}}}\ (\bibinfo {year} {2024})\BibitemShut {NoStop}%
\bibitem [{\citenamefont {Thavamani}\ \emph {et~al.}(2020)\citenamefont
  {Thavamani}, \citenamefont {Cuvier}, \citenamefont {Willert}, \citenamefont
  {Foucaut}, \citenamefont {Atkinson},\ and\ \citenamefont
  {Soria}}]{Thavamani2020}%
  \BibitemOpen
  \bibfield  {author} {\bibinfo {author} {\bibfnamefont {A.}~\bibnamefont
  {Thavamani}}, \bibinfo {author} {\bibfnamefont {C.}~\bibnamefont {Cuvier}},
  \bibinfo {author} {\bibfnamefont {C.}~\bibnamefont {Willert}}, \bibinfo
  {author} {\bibfnamefont {J.}~\bibnamefont {Foucaut}}, \bibinfo {author}
  {\bibfnamefont {C.}~\bibnamefont {Atkinson}},\ and\ \bibinfo {author}
  {\bibfnamefont {J.}~\bibnamefont {Soria}},\ }\href
  {https://doi.org/10.1016/j.expthermflusci.2020.110080} {\bibfield  {journal}
  {\bibinfo  {journal} {Experimental Thermal and Fluid Science}\ }\textbf
  {\bibinfo {volume} {115}},\ \bibinfo {pages} {110080} (\bibinfo {year}
  {2020})}\BibitemShut {NoStop}%
\bibitem [{\citenamefont {de~Silva}\ \emph {et~al.}(2016)\citenamefont
  {de~Silva}, \citenamefont {Hutchins},\ and\ \citenamefont
  {Marusic}}]{deSilva2016}%
  \BibitemOpen
  \bibfield  {author} {\bibinfo {author} {\bibfnamefont {C.~M.}\ \bibnamefont
  {de~Silva}}, \bibinfo {author} {\bibfnamefont {N.}~\bibnamefont {Hutchins}},\
  and\ \bibinfo {author} {\bibfnamefont {I.}~\bibnamefont {Marusic}},\ }\href
  {https://doi.org/10.1017/jfm.2015.672} {\bibfield  {journal} {\bibinfo
  {journal} {Journal of Fluid Mechanics}\ }\textbf {\bibinfo {volume} {786}},\
  \bibinfo {pages} {309} (\bibinfo {year} {2016})}\BibitemShut {NoStop}%
\bibitem [{\citenamefont {Sillero}\ \emph {et~al.}(2013)\citenamefont
  {Sillero}, \citenamefont {Jiménez},\ and\ \citenamefont
  {Moser}}]{Sillero2013}%
  \BibitemOpen
  \bibfield  {author} {\bibinfo {author} {\bibfnamefont {J.~A.}\ \bibnamefont
  {Sillero}}, \bibinfo {author} {\bibfnamefont {J.}~\bibnamefont {Jiménez}},\
  and\ \bibinfo {author} {\bibfnamefont {R.~D.}\ \bibnamefont {Moser}},\
  }\bibfield  {journal} {\bibinfo  {journal} {Physics of Fluids}\ }\textbf
  {\bibinfo {volume} {25}},\ \href {https://doi.org/10.1063/1.4823831}
  {10.1063/1.4823831} (\bibinfo {year} {2013})\BibitemShut {NoStop}%
\bibitem [{\citenamefont {Borrell}\ \emph {et~al.}(2013)\citenamefont
  {Borrell}, \citenamefont {Sillero},\ and\ \citenamefont
  {Jiménez}}]{Borrell2013}%
  \BibitemOpen
  \bibfield  {author} {\bibinfo {author} {\bibfnamefont {G.}~\bibnamefont
  {Borrell}}, \bibinfo {author} {\bibfnamefont {J.~A.}\ \bibnamefont
  {Sillero}},\ and\ \bibinfo {author} {\bibfnamefont {J.}~\bibnamefont
  {Jiménez}},\ }\href {https://doi.org/10.1016/j.compfluid.2012.07.004}
  {\bibfield  {journal} {\bibinfo  {journal} {Computers \& Fluids}\ }\textbf
  {\bibinfo {volume} {80}},\ \bibinfo {pages} {37} (\bibinfo {year}
  {2013})}\BibitemShut {NoStop}%
\bibitem [{\citenamefont {Simens}\ \emph {et~al.}(2009)\citenamefont {Simens},
  \citenamefont {Jiménez}, \citenamefont {Hoyas},\ and\ \citenamefont
  {Mizuno}}]{Simens2009}%
  \BibitemOpen
  \bibfield  {author} {\bibinfo {author} {\bibfnamefont {M.~P.}\ \bibnamefont
  {Simens}}, \bibinfo {author} {\bibfnamefont {J.}~\bibnamefont {Jiménez}},
  \bibinfo {author} {\bibfnamefont {S.}~\bibnamefont {Hoyas}},\ and\ \bibinfo
  {author} {\bibfnamefont {Y.}~\bibnamefont {Mizuno}},\ }\href
  {https://doi.org/10.1016/j.jcp.2009.02.031} {\bibfield  {journal} {\bibinfo
  {journal} {Journal of Computational Physics}\ }\textbf {\bibinfo {volume}
  {228}},\ \bibinfo {pages} {4218} (\bibinfo {year} {2009})}\BibitemShut
  {NoStop}%
\bibitem [{\citenamefont {Jiménez}(2013)}]{Javier2013}%
  \BibitemOpen
  \bibfield  {author} {\bibinfo {author} {\bibfnamefont {J.}~\bibnamefont
  {Jiménez}},\ }\bibfield  {journal} {\bibinfo  {journal} {Physics of Fluids}\
  }\textbf {\bibinfo {volume} {25}},\ \href {https://doi.org/10.1063/1.4824988}
  {10.1063/1.4824988} (\bibinfo {year} {2013})\BibitemShut {NoStop}%
\bibitem [{\citenamefont {Sillero}\ \emph {et~al.}(2014)\citenamefont
  {Sillero}, \citenamefont {Jiménez},\ and\ \citenamefont
  {Moser}}]{Sillero2014}%
  \BibitemOpen
  \bibfield  {author} {\bibinfo {author} {\bibfnamefont {J.~A.}\ \bibnamefont
  {Sillero}}, \bibinfo {author} {\bibfnamefont {J.}~\bibnamefont {Jiménez}},\
  and\ \bibinfo {author} {\bibfnamefont {R.~D.}\ \bibnamefont {Moser}},\
  }\bibfield  {journal} {\bibinfo  {journal} {Physics of Fluids}\ }\textbf
  {\bibinfo {volume} {26}},\ \href {https://doi.org/10.1063/1.4899259}
  {10.1063/1.4899259} (\bibinfo {year} {2014})\BibitemShut {NoStop}%
\bibitem [{\citenamefont {Jim\`{e}nez}\ \emph {et~al.}(2010)\citenamefont
  {Jim\`{e}nez}, \citenamefont {Hoyas}, \citenamefont {Simens},\ and\
  \citenamefont {Mizuno}}]{Jimenez2010}%
  \BibitemOpen
  \bibfield  {author} {\bibinfo {author} {\bibfnamefont {J.}~\bibnamefont
  {Jim\`{e}nez}}, \bibinfo {author} {\bibfnamefont {S.}~\bibnamefont {Hoyas}},
  \bibinfo {author} {\bibfnamefont {M.~P.}\ \bibnamefont {Simens}},\ and\
  \bibinfo {author} {\bibfnamefont {Y.}~\bibnamefont {Mizuno}},\ }\href
  {https://doi.org/10.1017/S0022112010001370} {\bibfield  {journal} {\bibinfo
  {journal} {Journal of Fluid Mechanics}\ }\textbf {\bibinfo {volume} {657}},\
  \bibinfo {pages} {335} (\bibinfo {year} {2010})}\BibitemShut {NoStop}%
\bibitem [{\citenamefont {Eisma}\ \emph {et~al.}(2015)\citenamefont {Eisma},
  \citenamefont {Westerweel}, \citenamefont {Ooms},\ and\ \citenamefont
  {Elsinga}}]{Eisma2015}%
  \BibitemOpen
  \bibfield  {author} {\bibinfo {author} {\bibfnamefont {J.}~\bibnamefont
  {Eisma}}, \bibinfo {author} {\bibfnamefont {J.}~\bibnamefont {Westerweel}},
  \bibinfo {author} {\bibfnamefont {G.}~\bibnamefont {Ooms}},\ and\ \bibinfo
  {author} {\bibfnamefont {G.~E.}\ \bibnamefont {Elsinga}},\ }\bibfield
  {journal} {\bibinfo  {journal} {Physics of Fluids}\ }\textbf {\bibinfo
  {volume} {27}},\ \href {https://doi.org/10.1063/1.4919909}
  {10.1063/1.4919909} (\bibinfo {year} {2015})\BibitemShut {NoStop}%
\bibitem [{\citenamefont {Chen}\ and\ \citenamefont
  {Blackwelder}(1978)}]{Chen1978}%
  \BibitemOpen
  \bibfield  {author} {\bibinfo {author} {\bibfnamefont {C.-H.~P.}\
  \bibnamefont {Chen}}\ and\ \bibinfo {author} {\bibfnamefont {R.~F.}\
  \bibnamefont {Blackwelder}},\ }\href
  {https://doi.org/10.1017/S0022112078002438} {\bibfield  {journal} {\bibinfo
  {journal} {Journal of Fluid Mechanics}\ }\textbf {\bibinfo {volume} {89}},\
  \bibinfo {pages} {1} (\bibinfo {year} {1978})}\BibitemShut {NoStop}%
\bibitem [{\citenamefont {Kovasznay}(1970)}]{Kovasznay1970}%
  \BibitemOpen
  \bibfield  {author} {\bibinfo {author} {\bibfnamefont {L.~S.~G.}\
  \bibnamefont {Kovasznay}},\ }\href
  {https://doi.org/10.1146/annurev.fl.02.010170.000523} {\bibfield  {journal}
  {\bibinfo  {journal} {Annual Review of Fluid Mechanics}\ }\textbf {\bibinfo
  {volume} {2}},\ \bibinfo {pages} {95} (\bibinfo {year} {1970})}\BibitemShut
  {NoStop}%
\bibitem [{\citenamefont {Hedley}\ and\ \citenamefont
  {Keffer}(1974)}]{Hedley1974}%
  \BibitemOpen
  \bibfield  {author} {\bibinfo {author} {\bibfnamefont {T.~B.}\ \bibnamefont
  {Hedley}}\ and\ \bibinfo {author} {\bibfnamefont {J.~F.}\ \bibnamefont
  {Keffer}},\ }\href {https://doi.org/10.1017/S0022112074001844} {\bibfield
  {journal} {\bibinfo  {journal} {Journal of Fluid Mechanics}\ }\textbf
  {\bibinfo {volume} {64}},\ \bibinfo {pages} {645} (\bibinfo {year}
  {1974})}\BibitemShut {NoStop}%
\bibitem [{\citenamefont {Corrsin}(1943)}]{Corrsin1943}%
  \BibitemOpen
  \bibfield  {author} {\bibinfo {author} {\bibfnamefont {S.}~\bibnamefont
  {Corrsin}},\ }\href@noop {} {\emph {\bibinfo {title} {Investigation of Flow
  in an Axially Symmetrical Heated Jet of Air}}},\ \bibinfo {type} {Tech.
  Rep.}\ \bibinfo {number} {NACA-ACR-3L23}\ (\bibinfo  {institution}
  {California Institute of Technology},\ \bibinfo {year} {1943})\BibitemShut
  {NoStop}%
\bibitem [{\citenamefont {Chauhan}\ \emph
  {et~al.}(2014{\natexlab{b}})\citenamefont {Chauhan}, \citenamefont {Philip},\
  and\ \citenamefont {Marusic}}]{Chauhan2014b}%
  \BibitemOpen
  \bibfield  {author} {\bibinfo {author} {\bibfnamefont {K.}~\bibnamefont
  {Chauhan}}, \bibinfo {author} {\bibfnamefont {J.}~\bibnamefont {Philip}},\
  and\ \bibinfo {author} {\bibfnamefont {I.}~\bibnamefont {Marusic}},\ }\href
  {https://doi.org/10.1017/jfm.2014.298} {\bibfield  {journal} {\bibinfo
  {journal} {Journal of Fluid Mechanics}\ }\textbf {\bibinfo {volume} {751}},\
  \bibinfo {pages} {298} (\bibinfo {year} {2014}{\natexlab{b}})}\BibitemShut
  {NoStop}%
\end{thebibliography}%

\end{document}